\documentclass[pra,aps,superscriptaddress,twocolumn,nofootinbib,longbibliography,accepted=2024-06-17]{quantumarticle}
\pdfoutput=1
\usepackage{amsfonts,amsmath,amssymb,amsthm,bm,bbm,centernot,color,xcolor,enumerate,enumitem,graphicx,hyperref,mathdots,mathtools,multirow,soul,subfigure,thmtools}
\usepackage[capitalise, noabbrev]{cleveref}
\hypersetup{colorlinks=true,linkcolor=blue,citecolor=blue,urlcolor=blue}
\usepackage[margin=2cm]{geometry}
\usepackage[utf8]{inputenc}
\usepackage[numbers]{natbib}

\definecolor{darkgreen}{RGB}{0,175,0}
\definecolor{ppblue}{RGB}{46,117,182}
\definecolor{ppred}{RGB}{197, 90, 17}

\let\LL\L

\def\C{ {\cal C} }
\def\Q{ {\cal Q} }
\def\E{ {\cal E} }
\def\F{ {\cal F} }
\def\I{ {\cal I} }
\def\L{ {\cal L} }
\def\T{ {\cal T} }
\def\U{ {\cal U} }
\def\V{ {\cal V} }

\def\R{ \mathbb{R} }

\def\>{\rangle}
\def\<{\langle}
\newcommand{\bra}[1]{\langle {#1} |}
\newcommand{\ket}[1]{| {#1} \rangle}
\newcommand{\abs}[1]{\left| {#1} \right|}
\newcommand{\ketbra}[2]{\ensuremath{\left|#1\right\rangle\!\!\left\langle#2\right|}}

\newcommand{\matrixel}[3]{\ensuremath{\left\langle #1 \vphantom{#2#3} \right| #2 \left| #3 \vphantom{#1#2} \right\rangle}}
\newcommand{\tr}[1]{\mathrm{Tr}\left( #1 \right)}

\newcommand{\iden}{\mathbbm{1}}

\theoremstyle{plain}
\newtheorem{thm}{Theorem}
\newtheorem{lem}[thm]{Lemma}
\newtheorem{prop}[thm]{Proposition}
\newtheorem{cor}[thm]{Corollary}

\theoremstyle{definition}

\newtheorem{defn}{Definition}

\newcommand{\proj}[1]     {\left| #1\middle\rangle\!\middle\langle#1\right|}
\newcommand{\Tr}          {\mathrm{Tr}}
\newcommand{\e}{{\mathrm e}}
\renewcommand{\cal}[1]{\mathcal{#1}}
\renewcommand{\v} {\mathbb{V}}

\newcommand{\Bra}[1]{\langle\!\langle {#1} |}
\newcommand{\Ket}[1]{| {#1} \rangle\!\rangle}

\begin{document}

\title{Quantum-embeddable stochastic matrices}

\author{Fereshte Shahbeigi}
\affiliation{Faculty of Physics, Astronomy and Applied Computer Science, Jagiellonian University, 30-348 Krakow, Poland}

\author{Christopher T.~Chubb}
\affiliation{Institute for Theoretical Physics, ETH Zürich, 8093 Zürich, Switzerland}

\author{Ryszard Kukulski}
\affiliation{Institute of Theoretical and Applied Informatics, Polish Academy of Sciences, Bałtycka 5, 44-100 Gliwice, Poland}
\affiliation{Faculty of Physics, Astronomy and Applied Computer Science, Jagiellonian University, 30-348 Krakow, Poland}

\author{\LL{}ukasz Pawela}
\affiliation{Institute of Theoretical and Applied Informatics, Polish Academy of Sciences, Bałtycka 5, 44-100 Gliwice, Poland}

\author{Kamil Korzekwa}
\affiliation{Faculty of Physics, Astronomy and Applied Computer Science, Jagiellonian University, 30-348 Krakow, Poland}
\date{\today}

\maketitle

\begin{abstract}
    The classical embeddability problem asks whether a given stochastic matrix $T$, describing transition probabilities of a $d$-level system, can arise from the underlying homogeneous continuous-time Markov process. Here, we investigate the quantum version of this problem, asking of the existence of a Markovian quantum channel generating state transitions described by a given $T$. More precisely, we aim at characterising the set of quantum-embeddable stochastic matrices that arise from memoryless continuous-time quantum evolution. To this end, we derive both outer and inner approximations on that set, providing new families of stochastic matrices that are quantum-embeddable but not classically-embeddable, as well as families of stochastic matrices that are not quantum-embeddable. As a result, we demonstrate that a larger set of transition matrices can be explained by memoryless models if the dynamics is allowed to be quantum, but we also identify a non-zero measure set of random processes that cannot be explained by either classical or quantum memoryless dynamics. Finally, we fully characterise extreme stochastic matrices (with entries given only by zeros and ones) that are quantum-embeddable.
\end{abstract}


\section{Introduction}

In 1937, a Finnish mathematician, Gustav Elfving, asked a fundamental question concerning the nature of randomness~\cite{elfving1937theorie}. Namely, he wondered which of the observed random transitions between discrete states of a given system can be explained by the underlying continuous memoryless process. More precisely, consider a system with $d$ distinguishable states and initially prepared in some state~$j$, which then evolves for a time~$t_f$, and after that the system is measured and found in some state~$i$. Repeating this experiment many times and recording the frequency of observed output states for all input states, one recovers a transition matrix $T$ with matrix elements~$T_{ij}$ describing the probabilities of state transitions from~$j$ to $i$. Elfving then asked, whether the random process observed at time~$t_f$ and described by~$T$ can arise from a homogeneous continuous-time Markov process, i.e., a process acting continuously and identically at all times $t\in[0,t_f]$, and such that the evolution of the system at each infinitesimal moment in time depends only on the current state of the system (and not on its history). Formally, this corresponds to the following \emph{embedding problem}~\cite{davies2010embeddable}: given a transition matrix~$T$, one wants to know whether there exists a family of transition matrices $e^{L t}$ continuously connecting the identity at $t=0$ with $T=e^{L t_f}$ at $t=t_f$. Here, $L$ is a time-independent \emph{Markov generator} (also known as the rate matrix) that is a $d\times d$ matrix with non-negative off-diagonal entries and columns summing to zero~\cite{davies2010embeddable}.

Despite decades of efforts, complete solutions to the classical embeddability problem described above have only been found for $2\times 2$~\cite{kingman1962imbedding}, $3\times 3$~\cite{cuthbert1973logarithm,johansen1974some,carette1995characterizations}, and very recently also for $4 \times 4$ matrices~\cite{casanellas2020embedding}. Nevertheless, the subset of embeddable matrices of general size $d$ can be bounded within the set of all $d\times d$ stochastic matrices, because various necessary conditions for embeddability were found, among them the following one~\cite{goodman1970intrinsic}:
\begin{equation}
    \label{eq:necessary}
    \prod_{i=1}^d T_{ii} \geq \mathrm{det~}T \geq 0.
\end{equation}
Thus, every continuous-time random process that results in a transition matrix $T$ failing to satisfy the above conditions must necessarily use memory effects\footnote{This holds even if we drop the time-homogeneity assumption, because Eq.~\eqref{eq:necessary} is a necessary condition for embeddability in the stronger sense, when the Markov generator is allowed to be time-dependent.}. In other words, observing such a random process $T$ only at a discrete moment in time, we can infer that the underlying dynamics cannot be explained with a memoryless model.

\begin{figure}
    \centering
    \includegraphics[width=\columnwidth]{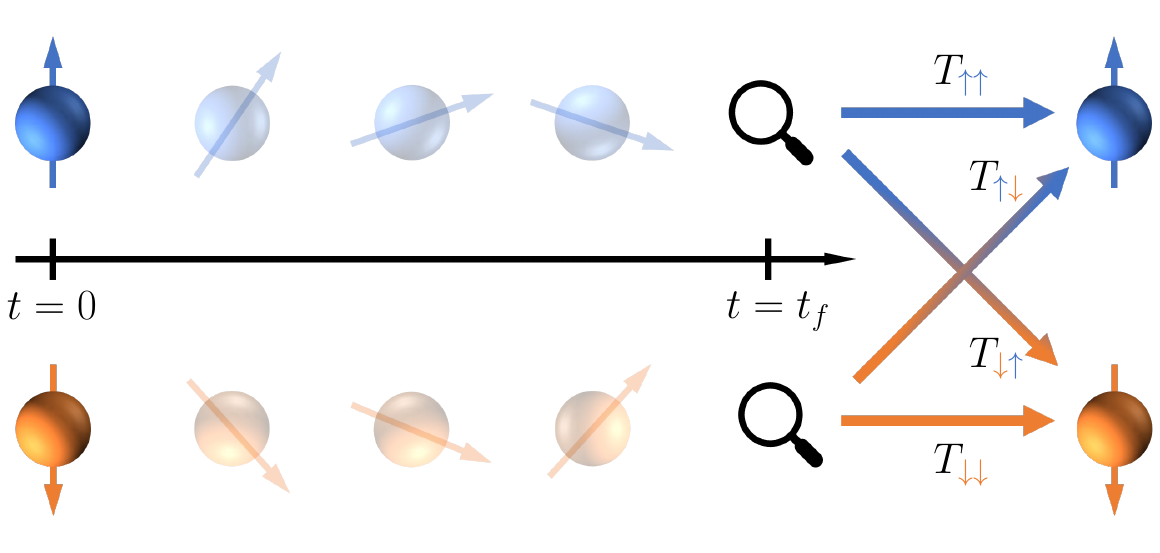}
    \caption{\label{fig:intro} \textbf{Transition matrices arising from memoryless quantum dynamics.} A two-level quantum system prepared initially in one of two perfectly distinguishable states, $\ket{\uparrow}$ or $\ket{\downarrow}$, undergoes an unknown evolution for time $t_f$ after which it is measured in the $\{\ket{\uparrow},\ket{\downarrow}\}$ basis. The conditional outcome probabilities, i.e., probabilities $T_{ij}$ of observing $\ket{i}$ given that the system started in $\ket{j}$ for $i,j\in\{\uparrow,\downarrow\}$, form a transition matrix~$T$. One can then ask which observed $T$ can be explained by the underlying time-homogeneous memoryless quantum dynamics. The central problem we investigate in this paper is the generalisation of this question to $d$-dimensional quantum systems.
    }
\end{figure}

In this paper we ask: how the non-existence of a memoryless model explaining a given $T$ is affected if we drop the implicit assumption that the underlying continuous dynamical process is classical and instead allow for a quantum evolution? This means that instead of occupying one of the well-defined $d$ states, the investigated system can be in any coherent superposition of these $d$ states, and the general evolution consists of both classical stochastic jumps and quantum coherent transitions. For the simplest example that we illustrate in Fig.~\ref{fig:intro}, consider a spin-1/2 particle that is initially prepared in one of the two perfectly distinguishable states, either a spin-up state $\ket{\uparrow}$ or a spin-down state $\ket{\downarrow}$. As in the classical scenario, it is then left to evolve for a time $t_f$, after which the orientation of its spin is measured, the procedure is repeated and the outcome frequencies yield the transition matrix~$T$. The crucial difference is that the evolution between time $0$ and $t_f$ is quantum, and we want to ask whether a given $T$ can arise from a memoryless quantum dynamics.

This problem of \emph{quantum-embeddability} of stochastic matrices was recently introduced in Ref.~\cite{korzekwa2021quantum}, where the authors focused on a more general concept of memoryless quantum dynamics that is time-inhomogeneous, i.e., the generator of the evolution may change in time. Here, in the spirit of the original Elfving problem, we want to characterise the set of transition matrices that can arise from the underlying homogeneous continuous-time quantum Markov process. Our main results consist of inner and outer approximations on the set of quantum-embeddable stochastic matrices, i.e., we identify both a new family of stochastic matrices that are quantum-embeddable (but not classically-embeddable) and a family of stochastic matrices that are not quantum-embeddable (neither classically-embeddable). As a result, we find a whole class of random processes that cannot arise from either a classical or quantum memoryless dynamics. Therefore, observing such a process only at a discrete moment in time, we can infer that the underlying dynamics is non-Markovian.

The paper is structured as follows. First, in Sec.~\ref{sec:setting}, we provide the mathematical background for our studies and formally state the investigated problem. Then, in Sec.~\ref{sec:results}, we present and discuss our main results concerning families of transition matrices that are not quantum-embeddable, and those that are. Section~\ref{sec:derivation} contains step-by-step derivations of our results together with intermediate results that may be of independent interest. Finally, Sec.~\ref{sec:conclusions} contains conclusions and an outlook for future work.

\section{Setting the scene}
\label{sec:setting}

A state of a $d$-dimensional quantum system is described by a $d\times d$ density matrix $\rho$ that is positive and has unit trace. A general open quantum evolution of such a system is given by a quantum channel $\E$, which is a linear map between $d\times d$ matrices that is completely positive and trace-preserving. A channel $\E$ is called time-independent embeddable (Markovian) if and only if it is in the closure of the maps of the following form~\cite{wolf2008assessing}:
\begin{equation}
	\label{eq:q_markovian}
	\E_t = e^{\L t},
\end{equation}
where $t\in\mathbb{R}_+$ denotes a finite time and $\L$ is the Lindblad generator satisfying~\cite{gorini1976completely,lindblad1976generators}
\begin{equation}
	\label{eq:lindbladian}
	\L(\cdot)=i[\cdot,H] + \Phi(\cdot)-\frac{1}{2}\{\Phi^*(\iden),\cdot\}.
\end{equation}
Above, $H$ is Hermitian and physically corresponds to the Hamiltonian of the system that induces the closed unitary dynamics, $\Phi$ is a completely positive (but not necessarily trace-preserving) map that physically describes dissipative open dynamics due to interactions with the environment, $\{\cdot,\cdot\}$ is the anticommutator and $\Phi^*$ denotes the dual of $\Phi$, i.e.,
\begin{equation}
\forall{A,B}:~~\tr{A \Phi(B)}=\tr{\Phi^*(A)B}.
\end{equation}
To be physically meaningful, $\L$ is assumed to have a finite operator norm. Thus, the assumption of closure is necessary in order to include in the set of Markovian channels the non-invertible channels that can be generated by Markovian dynamics with arbitrary precision.

A classical action $T$ of a quantum channel $\E$ is defined by a stochastic matrix describing transitions induced by~$\E$ in a given basis $\{\ket{i}\}_{i=1}^d$:
\begin{equation}\label{eq:E to T}
	T_{ij} := \matrixel{i}{\E(\ketbra{j}{j})}{i}.
\end{equation}
The set of all $d\times d$ stochastic matrices (i.e., matrices with non-negative entries and columns summing to identity) will be denoted by $\T_d$. Now, the central notion investigated in this paper is the following time-homogeneous version of quantum-embeddability~\cite{korzekwa2021quantum}.

\begin{defn}[Quantum-embeddable stochastic matrix]
    A stochastic matrix $T$ is \emph{quantum-embeddable} if it is a classical action of some Markovian quantum channel, i.e., if, for any $\delta>0$, there exist a Lindbladian $\L$ and a finite time $t_f$ such that
    \begin{equation}
    \label{eq:Embeddability_definition}
       \max_{i,j}\big|T_{ij}-\matrixel{i}{e^{\L t_f}(\ketbra{j}{j})}{i}\big|\leq\delta.
    \end{equation}
\end{defn}
The above definition generalises to the quantum setting the set of classical limit-embeddable stochastic matrices introduced in Ref.~\cite{wolpert2019}. It is worth mentioning that since the above definition includes all $\delta>0$, it is invariant for all the equivalent metrics. In particular, replacing the above definition with any distance induced by a matrix norm gives the same result.

We will denote the set of $d\times d$ quantum-embeddable stochastic matrices by $\Q_d$ and its complement within the set of $d\times d$ stochastic matrices by $\Q_d^c$. Similarly, the set of classically-embeddable stochastic matrices described in the introduction and its complement will be denoted by $\C_d$ and $\C_d^c$. Note that both sets $\C_d$ and $\Q_d$ are, by definition, closed. Moreover, it is clear that $\C_d\subset \Q_d$, since classical processes form a particular subset of quantum processes. This can be shown explicitly by noting that a classically-embeddable $T=e^{L t_f}$ is a classical action of a Markovian quantum channel $e^{\L t_f}$ with a Lindblad generator defined through Eq.~\eqref{eq:lindbladian} with $H=0$ and
\begin{equation}
    \Phi(\cdot)=\sum_{i,j=1}^d |L_{ij}|\ketbra{i}{j}(\cdot) \ketbra{j}{i}.
\end{equation}
On the other extreme, we can choose $\Phi=0$ and by varying $H$ generate all unitary channels. Thus, $\U_d\subset \Q_d$, with $\U_d$ denoting the set of $d\times d$ unistochastic matrices~\cite{bengtsson2004importance}, i.e., matrices $T$ such that $T_{ij}=|U_{ij}|^2$ for some unitary matrix $U$. Since there exist matrices in $\U_d$ but not in $\C_d$ (e.g., a non-trivial permutation matrix), and there exist matrices in $\C_d$ and not in $\U_d$ (e.g., a matrix $T$ with $T_{ij}=\delta_{ik}$ for any fixed $k$), $\Q_d$ is a strict superset of both $\U_d$ and $\C_d$. This approximates the set $\Q_d$ within the set $\T_d$ from inside, and the aim of this paper is to improve this bound, as well as to approximate $\Q_d$ from outside, i.e., to identify which stochastic matrices are not quantum-embeddable.


\section{Results and discussion}
\label{sec:results}

\begin{figure}
    \centering
    \includegraphics[width=0.48\columnwidth]{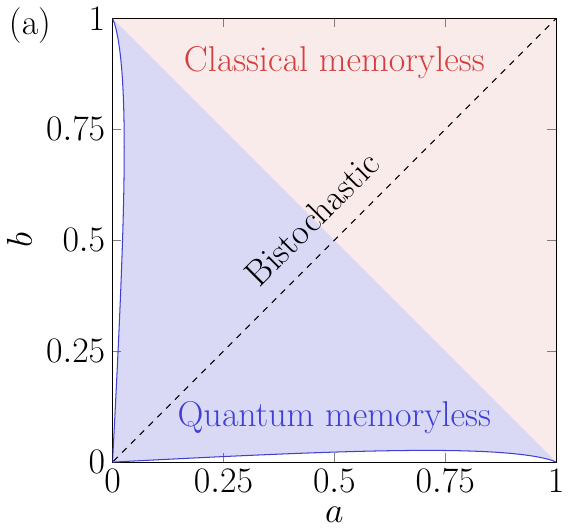}
	\includegraphics[width=0.48\columnwidth]{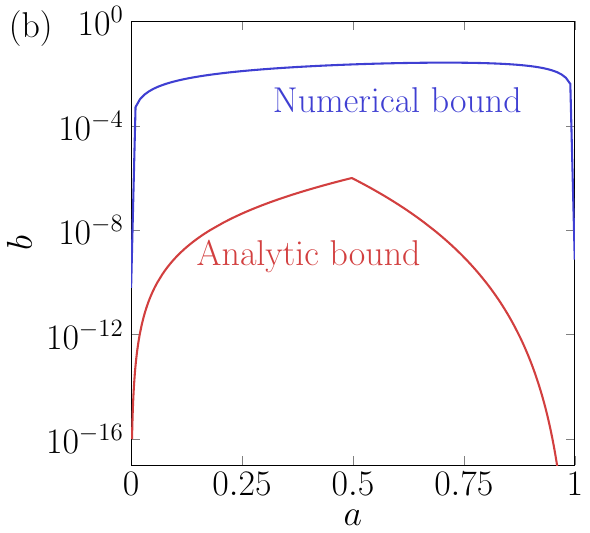}
    \caption{\label{fig:qubit_regions} \textbf{Memoryless dynamics of a two-level system.} Every $2\times 2$ stochastic matrix representing the evolution of a two-level system (see Eq.~\eqref{eq:2x2_stochastic}) can be represented as a point inside a square \mbox{$(a,b)\in[0,1]\times[0,1]$}. (a) The upper-right red region contains classically-embeddable matrices, whereas the bottom-left blue region contains those quantum-embeddable matrices that are not classically-embeddable. For transitions described by matrices belonging to the white region (left or below the blue bounding curves obtained through numerical optimisations) there are no memoryless models (classical or quantum) explaining them. (b)~Comparison of the analytical and numerical bound for the set $\Q_2$: above the top blue curve all matrices are quantum-embeddable, whereas below the bottom red curve no matrices are quantum-embeddable. The real boundary between $\Q_2$ and $\Q_2^c$ lies somewhere between the two curves.
    }
\end{figure}

Our first result identifies a subset of $2\times 2$ matrices that are not quantum-embeddable.
\begin{thm}
    \label{thm:qubit}
    Consider a $2\times 2$ stochastic matrix $T$,
    \begin{equation}
    \label{eq:2x2_stochastic}
        T=\begin{pmatrix}
            a&1-b\\
            1-a&b
        \end{pmatrix}.
    \end{equation}
    If $a\leq 10^{-6}$ and $f(a)\left(2-f(a)\right)< b < 1-g(a)$, then  $T\notin \Q_2$. Here,
    \begin{subequations}\label{eq:f and g_app}
    \begin{align}
    f(a)=&~2\sqrt{2}a^{0.25}+\sqrt{a(2-a)}+a^{0.9}\label{subeq:f}\\ \nonumber
    &~+\frac{0.01(4\sqrt{a}+a^{0.45})}{1-(8\sqrt{a}+a^{0.45})}+2\sqrt{8\sqrt{a}+a^{0.45}},\\
    g(a)=&~(2-a)(2a+a^{0.1}),\label{subeq:g}
    \end{align}
    \end{subequations}
    so that both these functions vanish when $a\to 0$. The same result holds if we swap $a$ for $b$.
\end{thm}
The proof of the above theorem can be found in Sec.~\ref{sec:qubit}, whereas in Sec.~\ref{sec:geometric} we present its simplified version that only works for a special case of $a=0$ (or $b=0$ after swapping $a$ for $b$). Here, based on numerical investigations, we note that the bounds on $a$ and $b$ from Theorem~\ref{thm:qubit} are quite loose and most probably can be significantly improved. More precisely, numerically optimising over all Lindblad evolutions of a qubit system (see Appendix~\ref{app:numerics} for details), we observe that the numerical bounds on $a$ and $b$ differ by more than four orders of magnitudes from the analytic bounds stated in Theorem~\ref{thm:qubit}. We illustrate this in Fig.~\ref{fig:qubit_regions}, where in panel (a) the set $\T_2$ is presented, together with its subsets~$\C_2$ (characterised analytically in Ref.~\cite{kingman1962imbedding}) and $\Q_2$ (characterised numerically here); whereas in panel (b) we compare the numerical and analytic bounds.

Theorem~\ref{thm:qubit} shows that $\Q_2^c$ occupies a non-zero volume within the set $\T_2$ and thus outer approximates $\Q_2$ (note that this is contrary to the case of time-dependent generators, where the entire $\T_2$ can be generated by time-inhomogeneous quantum Markovian dynamics~\cite{korzekwa2021quantum}). Our second result provides such an outer approximation for higher dimensional systems. Namely, the following theorem identifies a family of matrices belonging to $\Q_d^c$ and so, since $\Q_d^c$ is non-empty and open, it shows that $\Q_d^c$ has a non-zero measure, which in turn yields a superset of~$\Q_d$.

\begin{thm}
    \label{thm:main}
    Consider a $d\times d$ stochastic matrix $T$. Let $\I_0\subset\{1,\cdots,d\}$ be a subset of indices such that $T$ invariantly permutes  $\I_0$. Also, let $\I_1\subset\I_0^c$ denote a subset of  the complementary set of $\I_0$, where for any $i_1\in\I_1$ it holds that $T_{i_0i_1}=1$ for some fixed $i_0\in\I_0$. Then, $T\in\Q_d^c$ if there exists an index $i$ such that
    \begin{equation}
        \sum_{i_1\in\I_1}T_{i_1i}=1.
    \end{equation}
\end{thm}

\begin{figure}
    \centering
    \includegraphics[width=0.85\columnwidth]{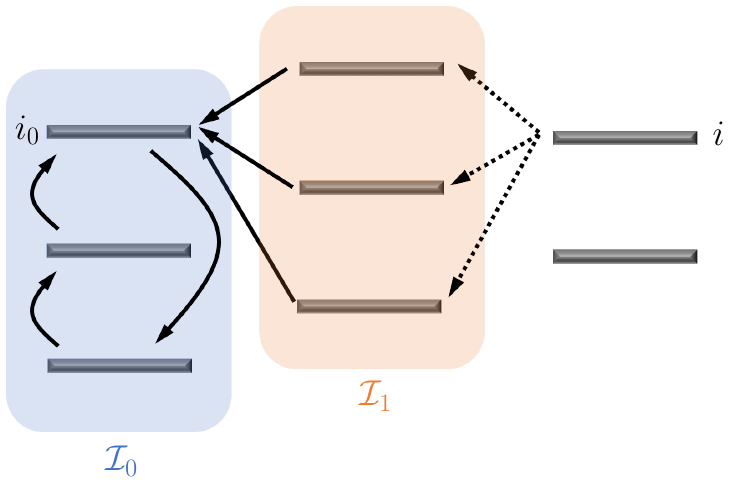}
	\caption{\label{fig:thm2} \textbf{Visualisation of Theorem~\ref{thm:main}.} An example of a transition matrix for a $d=8$ level system that is not quantum-embeddable according to Theorem~\ref{thm:main}. Solid arrows denote deterministic transitions (with probability 1), dotted arrows denote probabilistic transitions, and a lack of arrows from a given level means that arbitrary transitions from that level are allowed.
    }
\end{figure}

The proof of the above theorem can be found in Sec.~\ref{sec:qudit_nonembeddable}, whereas in Fig.~\ref{fig:thm2} we illustrate the sets, states and transitions appearing in the statement of the theorem to better visualise its content. Here, we will unpack Theorem~\ref{thm:main} by examining the lowest possible dimensions of $d=3$ and $d=4$. For $d=3$, the structure imposed by the theorem requires $\I_0$ to be one-dimensional. Thus, $T$ acts trivially on $\I_0=\{i_0\}$, i.e., $T_{i_0i_0}=1$. Moreover, $\I_1$ also needs to be one-dimensional. For different choices of $i_0$ and $i_1$, we then obtain that the following six extreme stochastic matrices are \emph{not} quantum-embeddable:
\begin{equation}
\begin{aligned}
    \label{eq:non_embeddable_3}
    &\left\{\begin{pmatrix}
        1&1&0\\
        0&0&1\\
        0&0&0
    \end{pmatrix},
    \begin{pmatrix}
        1&0&1\\
        0&0&0\\
        0&1&0
    \end{pmatrix},
    \begin{pmatrix}
        0&0&1\\
        1&1&0\\
        0&0&0
    \end{pmatrix},\right.\\
    &\left.~~\begin{pmatrix}
        0&0&0\\
        0&1&1\\
        1&0&0
    \end{pmatrix},
    \begin{pmatrix}
        0&1&0\\
        0&0&0\\
        1&0&1
    \end{pmatrix},
    \begin{pmatrix}
        0&0&0\\
        1&0&0\\
        0&1&1
    \end{pmatrix}\right\}.
\end{aligned}
\end{equation}
Therefore, no $3\times 3$ transition matrices in the vicinity of the above matrices can result from time-homogeneous Markovian quantum dynamics. Interestingly, the above set becomes quantum-embeddable if we use a broader definition of embeddability allowing for time-inhomogeneous quantum dynamics. To see that, we note that a concatenation of Markovian quantum channels is Markovian with a time-dependent Lindbladian~\cite{Wolf-Cirac-dividing}. Now, assume $\E_1$ to be a channel that sends $\proj1$ to $\proj0$ and keeps invariant the other two basis states. Furthermore, let $\E_2$ be a quantum channel that sends $\proj2$ to $\proj1$ and acts trivially on the other two basis states. It is easy to see $\E_1$ and $\E_2$ are time-homogenous Markovian. Their concatenation $\E_2\circ\E_1$ is, therefore, a time-inhomogenous Markovian channel with the classical action as the first matrix in Eq.~\eqref{eq:non_embeddable_3}. However, we note that the existence of quantum non-embeddable maps in $d=3$, even if we allow for the time-inhomogenous dynamics, has already been discussed in Ref.~\cite{korzekwa2021quantum}.

Taking $d=4$ brings more freedom to construct not only a discrete set of extreme stochastic matrices belonging to $\Q_4^c$, but also continuous families of stochastic matrices in $\Q_4^c$ located at the boundaries of $\T_4$. For example, let us fix $\I_0=\{1\}$ and $\I_1=\{m,n\}$ for different choices of $m,n\in\{2,3,4\}$. Then, Theorem~\ref{thm:main} yields the following three families of matrices in $\Q_4^c$:
\begin{equation}
\begingroup 
\setlength\arraycolsep{4pt}
    \left\{\!\begin{pmatrix}
        1&1&1&0\\
        0&0&0&p\\
        0&0&0&q\\
        0&0&0&0
    \end{pmatrix},\!
    \begin{pmatrix}
         1&1&0&1\\
        0&0&p&0\\
        0&0&0&0\\
        0&0&q&0
    \end{pmatrix},\!
    \begin{pmatrix}
         1&0&1&1\\
        0&0&0&0\\
        0&p&0&0\\
        0&q&0&0
    \end{pmatrix}\!\right\},
    \endgroup
\end{equation}
where $p,q$ are non-negative with $p+q=1$.

Our third result provides a novel inner approximation on~$\Q_d$, i.e., it gives constructions of matrices in $\Q_d$ that were not previously known to belong to $\Q_d$. We already mentioned that both $\C_d$ and $\U_d$, the sets of classically-embeddable and unistochastic matrices, form the subsets of $\Q_d$. Clearly, if $T$ can be written as \mbox{$T=R_1\oplus R_2$} such that $R_1\in\C_{d_1}$ and $R_2\in\U_{d_2}$, then $T$ is a quantum-embeddable matrix which does not necessarily lie in $\C_d$ or~$\U_d$. Moreover, knowing the set of quantum-embeddable stochastic matrices of lower dimensions $d_1$ and $d_2$, their direct sum gives a quantum-embeddable matrix in a higher dimension \mbox{$d=d_1+d_2$}. However, as the following theorem shows, one can also non-trivially apply the knowledge of lower-dimensional quantum-embeddable matrices to construct higher-dimensional matrices in $\Q_d$.

 \begin{thm}
    \label{thm:embeddable}
    Let $R\in\Q_{d'}$ reside in a diagonal block of a stochastic matrix $T$ of dimension $d\geq d'$. Then, $T$ is quantum-embeddable if its columns outside those occupied by $R$ are copies of the columns occupied by $R$.
\end{thm}

We present a constructive proof of the above theorem in Sec.~\ref{sec:qudit_embeddable}. In order to discuss its scope here, without loss of generality we restrict ourselves to the case where $R$ lives in the corner of $T$. Thus, $T$ has the following block form:
\begin{equation}
\label{eq:block_form_matrix}
    T=\left(\begin{array}{@{}c|c@{}}
  \begin{matrix}
     R
  \end{matrix}
  & S \\
\hline
   0 &
  \begin{matrix}
   B
  \end{matrix}
\end{array}\right),
\end{equation}
where $R\in\Q_{d'}$ and $0$ denotes a rectangular matrix of zeros (which must be the case because $R$ is by definition stochastic). Theorem \ref{thm:embeddable} states that if each column of $S$ is a copy of some column of $R$ (so $B=0$), then $T$ is quantum-embeddable. While this theorem only yields a subset of quantum-embeddable matrices lying on the boundaries of the set of stochastic matrices, when one restricts to extreme points (i.e., stochastic matrices with entries equal to either 0 or 1) then it can be shown that Theorems~\ref{thm:main} and \ref{thm:embeddable} characterise two complementary sets. This means that a given extreme stochastic matrix is either not quantum-embeddable structured according to Theorem~\ref{thm:main}, or it is quantum-embeddable and of the form given by Theorem~\ref{thm:embeddable}. This result is summarised in the following corollary.

\begin{cor}
\label{cor:extreme_points}
An extreme stochastic matrix $T$ of size $d$ is quantum-embeddable if and only if it includes a permutation as a diagonal block, and its other columns are given by copies of the columns of this permutation.
\end{cor}

The proof of the above corollary is presented in Sec.~\ref{sec:extreme_points}. Knowing the necessary and sufficient condition for an extreme stochastic matrix to be quantum-embeddable, one can investigate how their number changes with the dimension. For $d=2$, all four extreme points (two permutations and two non-trivial classically-embeddable matrices sending both states to either 0 or 1) are quantum-embeddable. For $d=3$, among the $27$ extreme stochastic matrices the $6$ matrices specified by Theorem~\ref{thm:main} and given in Eq.~\eqref{eq:non_embeddable_3} are not quantum-embeddable. The remaining ones include $6$ unistochastic matrices (permutations), $9$ classically-embeddable matrices ($3$ completely contractive maps on the whole space and $6$ maps completely contractive on a two-dimensional subspace with a fixed orthogonal subspace), and $6$ quantum-embeddable matrices with the structure given by Theorem~\ref{thm:embeddable}, which are neither in $\C_3$ nor $\U_3$. Generally, for dimension $d$, the number $n(d)$ of quantum-embeddable matrices, out of $d^d$ extreme points, is given by
\begin{equation}
    \!\!\!n(d):=\sum_{m=1}^d\binom{d}{m}m!\ m^{d-m}=\sum_{m=1}^d\frac{d!}{(d-m)!}m^{d-m},
\end{equation}
where, according to Corollary~\ref{cor:extreme_points}, one counts all the ways of inserting an $m\times m$ permutation matrix (with binomial factor counting all possible placements inside a matrix of size $d$ and $m!$ accounting for different permutation matrices) times the number of choices for the remaining $(d-m)$ columns (each of them must be a copy of one of the $m$ columns of the permutation matrix, hence $m^{d-m}$ choices). As a result, the ratio of the number of extreme and not quantum-embeddable stochastic matrices to all extreme stochastic matrices, $1-n(d)/d^d$,  approaches one as $d$ increases.


\section{Derivation of results}
\label{sec:derivation}

In this section, we will first present a simplified proof of Theorem~\ref{thm:qubit}, which works only for a special case of $a=0$ (or $b=0$). This will serve as an illustration of the main ideas and intuitions behind the full proof of Theorem~\ref{thm:qubit} that follows. In fact, we will prove a slightly stronger result concerning qubit channels that are Markovian. We will then proceed to proving Theorems~\ref{thm:main} and \ref{thm:embeddable} for higher-dimensional quantum systems. Finally, we will explain how Corollary~\ref{cor:extreme_points} follows from these theorems.


\subsection{Simplified proof of Theorem~\ref{thm:qubit}}
\label{sec:geometric}

We start by noting that a Markovian channel is infinitely divisible \cite{Davalos2019divisibility}. By definition, an infinitely divisible channel can be written as
\begin{equation}
\label{eq:infinitely_divisible}
    \E=(\E^{1/n})^{n},
\end{equation}
for any $n\in\mathbb{N}$ with $\E^{1/n}$ being a quantum channel itself. Furthermore, to prove our point we need the following two lemmas.
\begin{lem}[Theorem~4.9 of Ref.~\cite{braun2014universal}]
	\label{lem:pure_image}
	The image of a qubit channel $\E$ contains zero, one, two or all pure states. In the last case, $\E$ is a unitary channel.
\end{lem}
\begin{lem}
    \label{lem:contraction}
    If a qubit channel sends two distinct pure states into a single pure state $\proj\psi$, then it sends all states to $\proj\psi$.
\end{lem}
\begin{proof}
   Note that if two distinct pure states are mapped to the same pure state $\proj\psi$, then their convex combination, which is a full rank state, is also mapped to $\proj{\psi}$. It is, however, known that a full rank state is sent by a channel to a pure state if and only if the entire state space is mapped to that pure state~\cite{braun2014universal}.
\end{proof}
We will now prove a result that directly yields a version of Theorem~\ref{thm:qubit}
restricted to $2\times 2$ stochastic matrices lying on the boundaries of the set $\T_2$, i.e., such that $a=0$ and $0<b<1$, or $b=0$ and $0<a<1$.
\begin{prop}
    \label{prop:qubit_simp}
	If a pure state $\proj{\psi}$ exists in the image of a Markovian qubit channel $\E$, then one of the following holds:
	\begin{enumerate}
	    \item\label{case:unitary} $\E$ is unitary.
	    \item\label{case:damping} $\E$ is a non-unital map with $\ket\psi$ being its fixed point.
	    \item\label{case:dephasing} $\E$ is dephasing with respect to the basis $\{\ket{\psi},\ket{\psi^\perp}\}$, implying that both of them are fixed points of the channel.
	\end{enumerate}
\end{prop}

Before proceeding with the proof, we emphasise that since non-unital qubit channels have only one fixed state, the second case above implies that the image of $\E^n$ converges to $\proj{\psi}$ as $n$ increases.

\begin{proof}
    Without loss of generality, let $\ket0$ be the pure state that is mapped to $\ket{\psi}$, i.e, $\E(\proj0)=\proj\psi$. The proof consists of two separate parts. First, we will assume that $\ket\psi\neq\ket0$ and show that either Case~\ref{case:unitary} or Case~\ref{case:damping} holds, with the latter possible only for a maximally contractive~$\E$, i.e., \mbox{$\E(\rho)=\proj\psi$} for all $\rho$. Then, we will consider $\ket\psi=\ket0$ and prove that only Cases~\ref{case:damping}~and~\ref{case:dephasing} are possible.

    Assume $\ket\psi\neq\ket0$ and note that, since $\E$ is Markovian, Eq.~\eqref{eq:infinitely_divisible} holds for it. In particular, take $n=3$ and for \mbox{$i\in\{1,2,3\}$} let $\rho_i:=\E^{i/3}(\proj0)$ be the image of $\proj{0}$ when acted on by the cube root map once, twice, or thrice. Now, assume it is not the case that $\rho_1$ and $\rho_2$ are both pure. This means that for at least one $i\in\{1,2\}$ the map $\E^{(3-i)/3}$ has to send a full rank state to a pure one, thus it is maximally contractive as proved in Ref.~\cite{braun2014universal}. Given infinite divisibility of~$\E$, this implies that the maps $\E^{1/3}$ and $\E^{2/3}$ do not have any other point but $\proj{\psi}$ in their image, contradicting the assumption that $\rho_1$ and $\rho_2$ are not both pure. Thus, it is only possible to have $\rho_i=:\proj{\psi_i}$ pure for any $i$.

    In what follows, we will discuss all possible scenarios that may happen depending on the distinctness of the states $\proj{\psi_i}$.
    \begin{enumerate}
        \item If all $\proj{\psi_i}$'s are distinct, then the cube root map $\E^{1/3}$ has three pure states in its image, implying it is a unitary map by Lemma~\ref{lem:pure_image}. Therefore, $\E$ is also unitary.

        \item If two successive states are the same, i.e., if  $\proj{\psi_1}=\proj{\psi_2}$ or $\proj{\psi_2}=\proj{\psi_3}$, then $\proj{\psi_3}=\proj{\psi}$ is a fixed point of $\E^{1/3}$ and, consequently, of $\E$. Since $\E$ sends both $\proj{0}$ and $\proj{\psi}$ to the latter, it is maximally contractive by Lemma~\ref{lem:contraction}. Note that this also means that $\E^{1/3}$ is maximally contractive and all the $\proj{\psi_i}$'s are equal.

        \item Finally, we prove by contradiction that it is not possible to have $\proj{\psi_1}=\proj{\psi_3}$ with $\proj{\psi_2}$ being distinct. First, note that since
        \begin{equation}
            \E^{1/3}(\proj0)=\E^{1/3}(\proj{\psi_2})=\proj{\psi_3},
        \end{equation}
        we need $\proj{\psi_2}=\proj{0}$, as otherwise two distinct pure states would be mapped to the same pure state and, by Lemma~\ref{lem:contraction}, this would mean all $\proj{\psi_i}$'s are equal, contradicting that $\proj{\psi_2}$ is distinct. This, however, means that
        \begin{subequations}
        \begin{align}
            \E^{2/3}(\proj0)&=\proj0,\\
            \E^{2/3}(\proj\psi)&=\proj\psi.
        \end{align}
        \end{subequations}
        Having two distinct fixed states, $\E^{2/3}$ has to be unital which, in turn, implies that its fixed pure states are orthogonal, i.e., $\proj\psi=\proj1$. This means $\E^{2/3}$ is a dephasing map and any power of a dephasing map is also a dephasing map. Consequently, $\E$ is dephasing with $\proj0$ as its fixed state, and thus it cannot send it to a distinct state $\proj{\psi}$, leading to a contradiction.
    \end{enumerate}
    With the above discussion, we conclude that if \mbox{$\proj{\psi}\neq\proj{0}$}, then $\E$ is either unitary or completely contractive to $\proj\psi$, which completes the first part of the proof.

    For the second part, assume \mbox{$\proj0=\proj{\psi}$}. Thus  $\E(\proj{\psi})=\proj{\psi}$ means that $\proj{\psi}$ is a fixed point of~$\E$. If $\E$ is non-unital, we already have Case~\ref{case:damping}. Otherwise, a unital channel $\E$ with a fixed point $\proj{\psi}$ has to have $\ket{\psi^{\perp}}\!\bra{\psi^{\perp}}$ as a fixed point as well. This proves that $\E$ is dephasing in the basis $\{\ket{\psi},\ket{\psi^\perp}\}$ and completes the proof.
\end{proof}

The above proposition enables us to directly show that all stochastic matrices in Fig.~\ref{fig:qubit_regions} on the boundaries with $a=0$ and $b=0$ (excluding the end points) belong to $\Q_2^c$. To see that, let $a=0$ and note that such a stochastic matrix is a classical action of some quantum channel $\E$ that sends $\ket0$ to $\ket1$, i.e., $\E(\proj0)=\proj1$. Since $\ket0$ is then not a fixed point of $\E$, Proposition~\ref{prop:qubit_simp} tells us that for~$\E$ to be Markovian, it has to be either a unitary channel or a maximally contractive channel into $\proj1$. A unitary map sending $\ket0$ to $\ket1$ has to send the latter state to the former one, meaning that $b=0$. On the other hand, if $\E$ is maximally contractive, then it sends $\ket1$ to itself, and thus $b=1$. The proof for the boundary with $b=0$ is analogous.

While the above reasoning already shows that there exist some stochastic matrices that are not quantum-embeddable for $d=2$ (which, given $\Q_2^c$ is an open set, proves that $\Q_2^c$ is not of measure zero), one wonders how far away from the boundaries can such non-embeddable matrices exist, i.e., how big the deviation of $a$ or $b$ from zero can be to still get matrices that are not quantum-embeddable. Intuitively, it is expected that for a sufficiently small deviation, a Markovian $\E$ should be close to either a unitary map or a completely contractive one. In what follows, we prove this intuition.


\subsection{Proof of Theorem~\ref{thm:qubit}}
\label{sec:qubit}

In order to prove Theorem~\ref{thm:qubit}, we start by introducing the notation for trace distance,
\begin{equation}
    D\left(\rho,\sigma\right):=\frac12\|\rho-\sigma\|_1,
\end{equation}
and its upper and lower bounds~\cite{FG99,puchala_distance_bound}
\begin{equation}\label{eq:fidelity-distance_app}
    1\!-\!\Tr\left(\rho\sigma\right)\!-\!M(\rho)M(\sigma)\!\leq D\left(\rho,\sigma\right)\leq\sqrt{1-F(\rho,\sigma)},
\end{equation}
where $M(\rho)=\sqrt{1-\Tr\rho^2}$, and $F(\rho,\sigma)=(\Tr|\sqrt{\rho}\sqrt{\sigma}|)^2$ denotes the Uhlmann fidelity.
Additionally, we will denote by
\begin{equation}
\label{eq:simple_notation}
    \rho_\psi(t):=\e^{\L t}(\proj\psi)=\e^{\L t}(\psi)
\end{equation}
the evolved state of $\psi:=\proj{\psi}$ after time $t$ under Lindbladian $\L$.
Furthermore, we define an $\epsilon$-purity-preserving map as follows.
\begin{defn}\label{def:epsilon-purity}
    A dynamical map $\e^{\L t}$ is called $\epsilon$\emph{-purity-preserving} on a state $\ket\psi$ for some time $t_f$ if there exists a pure state $\ket{\psi_t}$ for all $t\in[0,t_f]$ such that
     \begin{equation}\label{eq:epsilon-purity}
        D\left(\rho_\psi(t),\proj{\psi_t}\right)\leq\epsilon.
    \end{equation}
\end{defn}
Finally, the following lemma will assist us in proving our main point in this section.
\begin{lem}\label{lem:trajectory}
Assume that a dynamical map $\e^{\L t}$ is $\epsilon$-purity-preserving on a state $\psi$ for some time $t_f$, and denote by $\psi_t$ a pure state in $\epsilon$-neighbourhood of the evolved $\psi$ at each moment of time. Then, for all $ t_1,t_2$ satisfying \mbox{$t_1+t_2\leq t_f$}, the largest eigenvalue $\lambda$ of \mbox{$\rho_{\psi_{t_1}}\!(t_2)=\e^{\L t_2}(\psi_{t_1})$} is restricted to
\begin{equation}
    \lambda\left(\rho_{\psi_{t_1}}\!(t_2)\right)\geq1-2\epsilon.
\end{equation}
\end{lem}
\begin{proof}
    Note that for, any $t_1+t_2\leq t_f$, we have
    \begin{align}\label{eq:trajectory}
      &D\left(\rho_{\psi_{t_1}}\!(t_2),\psi_{t_1+t_2}\right)\nonumber\\
       &\quad \leq  D\left(\rho_{\psi_{t_1}}(t_2),\rho_{\psi}(t_1+t_2)\right)+D\left(\rho_{\psi}(t_1+t_2),\psi_{t_1+t_2}\right)\nonumber\\
       &\quad \leq D\left(\psi_{t_1},\rho_{\psi}(t_1)\right)+\epsilon\leq2\epsilon,
    \end{align}
    where we first used the triangle inequality, and then the data processing inequality, together with the assumption of $\epsilon$-purity preserving property. The proof is completed by using the fact that, for any state $\rho$, the largest eigenvalue $\lambda$ satisfies
    \begin{equation}
        1-\lambda=\min_{\ket\xi} D\left(\rho,\proj\xi\right),
    \end{equation}
    where the minimum is over all pure states $\ket\xi$.
\end{proof}

We will now present and prove the main technical result of this section, from which the proof of Theorem~\ref{thm:qubit} follows almost immediately. Informally, it states that if a dynamical map sends $\ket{0}$ approximately to $\ket{1}$, then either $\ket{1}$ goes approximately to $\ket{0}$, or every state goes approximately to $\ket{1}$.

\begin{thm}\label{thm:Qubit-General_app}
    Consider a qubit dynamical map $\e^{\L t}$ that, at some time $t_f$, sends the state $\ket 0$ to the $\epsilon$-neighbourhood of its orthogonal state $\ket{1}$, i.e.,
    \begin{equation}\label{eq:e-neighbour}
        D\left(\e^{\L t_f}\left(\proj0\right),\proj0\right)\geq1-\epsilon.
    \end{equation}
    Then, for any $\epsilon\leq 10^{-6}$, one of the following two inequalities hold:
    \begin{subequations}
    \begin{align}
        \label{item:uniatry_app}
            D\left(\e^{\L t_f}\left(\proj1\right),\proj1\right)&\geq 1-f(\epsilon),\\
            D\left(\e^{\L t_f}\left(\proj\xi\right),\proj1\right)&\leq \sqrt{g(\epsilon)},\label{item:AD_app}
    \end{align}
    \end{subequations}
    where $\ket{\xi}$ is any pure state, and $f$ and $g$ are the functions specified by Eqs.~\eqref{subeq:f}-\eqref{subeq:g}.
    Moreover, the equalities hold only if $\epsilon=0$.
\end{thm}
\begin{proof}
     Let the spectral decomposition of the state $\ket 0$ evolved under $e^{\L t}$ at each $t\in[0,t_f]$ be given by
    \begin{align}
        \label{eq:EvolutionOf0}
       \!\!\!\!\rho_0(t)=\lambda_t\proj{\psi_0(t)}+\left(1\!-\!\lambda_t\right)\proj{\psi_1(t)}\!,
    \end{align}
    such that $\lambda_t\geq1/2$ denotes the largest eigenvalue.  The proof will consist of two parts. First, we will assume
    \begin{equation}\label{eq:lambda_app}
        \!\!\forall t\in[0,t_f]: D\left(\rho_0(t),\proj{\psi_0(t)}\right)=1-\lambda_t\leq \epsilon^{0.9},
    \end{equation}
    and prove it implies that the dynamics is almost unitary which results in Eq.~\eqref{item:uniatry_app}. Then, we will show when this assumption does not hold, the dynamics is almost a full contraction and we obtain Eq.~\eqref{item:AD_app}.

    \textbf{ Almost unitary dynamics.} As proved in Lemma~\ref{lem:trajectory}, the $\epsilon^{0.9}$-purity-preserving assumption in Eq.~\eqref{eq:lambda_app} implies that for any $t,t'\in[0,t_f/2]$, the evolution $\e^{\L t'}$ is $2\epsilon^{0.9}$-purity-preserving on a state $\ket{\psi_0(t)}$.  In what follows, we will employ the Stinespring dilation of $\e^{\L t'}$ on states $\ket0$, $\ket1$, and $\ket{\psi_0(t)}$ to prove the first part. Note that, in the Stinespring picture, to realise a quantum channel acting on a $d$-dimensional system, it is enough to take the environment of dimension~$d^2$. Therefore, we will restrict ourselves to the environments of dimension $4$.

    Denote by $U(t')$ the unitary operator in a Stinespring dilation of the map $\e^{\L t'}$, so that
    \begin{subequations}\label{eq:PsiStin}
    \begin{align}
        \label{subeq:evolved0}
        \ket{\Psi_0(t')}:=&\ U(t')\ket{0}\ket{0}=\sqrt{\lambda_{t'}}\ket{\psi_0(t')}\ket{r_0(t')}
        \\ \nonumber
         &\qquad\qquad\qquad+\sqrt{1-\lambda_{t'}}\ket{\psi_1(t')}\ket{r_1(t')},\\
        \ket{\Psi_1(t')}:=&\ U(t')\ket{1}\ket{0}\label{subeq:evolved1}
        \\ \nonumber
        =&\ \sum_{i=0}^1\sum_{j=0}^3s_{ij}(t')\ket{\psi_i(t')}\ket{r_j(t')},\\
        \ket{\Psi_{\!\psi_0\!(t)}(t')}:=&\ U(t')\ket{\psi_0(t)}\ket{0}\label{subeq:evolved0+1}
          \\ \nonumber
        =&\ \sum_{i=0}^1\sum_{j=0}^3f_{ij}(t,t')\ket{\psi_i(t')}\ket{r_j(t')},
    \end{align}
    \end{subequations}
    where all the states are written in the Schmidt basis of~$\ket{\Psi_0(t')}$. Note that for $m=\{0,1,\psi_0(t)\}$,
    \begin{align}\label{eq:marginals}
           \!\!\!\! \rho_m(t')&\!=\!\e^{\L t'}\!\!(\proj m)\!=\!\Tr_2\!\left(\ket{\Psi_m(t')}\bra{\Psi_m(t')}\right).
    \end{align}

    The $\epsilon^{0.9}$-purity-preserving assumption leads to almost pure marginals, therefore, bounds the entanglement of the states $\ket{\Psi_0(t')}$ and $\ket{\Psi_{\!\psi_0\!(t)}(t')}$ for any $t,t'$ in $[0,t_f/2]$. We show in Appendix~\ref{app:s10} that this implies that $|s_{10}|$ becomes the dominant coefficient. More precisely, we prove in Appendix~\ref{app:s10} that for any $\epsilon\leq10^{-6}$ we have
    \begin{subequations}\label{eq:coefficients}
        \begin{align}
        \label{eq:s00_main}|s_{00}(t')|&\leq\sqrt{\epsilon^{0.9}},\\
        \label{eq:s10bound_main}|s_{10}(t')|&\geq 1-4\sqrt{\epsilon}.
        \end{align}
    \end{subequations}
    Moreover, the equality holds in the second equation above only if $\epsilon=0$. Since $s_{10}(t')$ is the coefficient of $\ket{\psi_1(t')}\ket{r_0(t')}$, being the leading coefficient it implies that any superposition of $\ket{\Psi_1(t')}$ and $\ket{\Psi_0(t')}$, with the highest projection along $\ket{\psi_0(t')}\ket{r_0(t')}$, is almost a product state as $\ket{r_0(t')}$ can be factorised. This, in turn, has two consequences. First, any pure state remains almost pure under $\e^{\L t'}$ at any $t'\leq t_f/2$ for sufficiently small $\epsilon$. Second, the evolution is also $\epsilon$-orthogonality-preserving, i.e., it preserves the orthogonality of any two initially orthogonal states, up to a function of $\epsilon$ which we introduce in the following.

    Strictly speaking, we prove in Appendix~\ref{app:orthogonality-purity} that, restricted to $\epsilon\leq10^{-6}$ and $t'\leq t_f/2$, and applying the map $\e^{\L t'}$ on two initially pure orthogonal states $\ket\upsilon=b\ket0+b_\perp\ket1$ and $\ket{\upsilon^\perp}=b^\ast_\perp\ket0-b^\ast\ket1$, one gets
        \begin{align}
    \label{eq:ortho-preserving}
        D\left(\rho_{\upsilon}(t'),\rho_{\upsilon^\perp}(t')\right)&\geq \bigg(1-\frac{0.01(4\sqrt{\epsilon}+\epsilon^{0.45})}{1-(8\sqrt{\epsilon}+\epsilon^{0.45})}\nonumber\\
        &-2\sqrt{8\sqrt{\epsilon}+\epsilon^{0.45}}\bigg)=:\ 1-h(\epsilon).
    \end{align}
    Moreover, the largest eigenvalue $\eta$ of $\rho_{\upsilon}(t')$ for any pure state $\ket\upsilon$ respects
    \begin{align}\label{eq:purity-proof}
        \eta\geq1-(8\sqrt{\epsilon}+\epsilon^{0.45}).
    \end{align}

    The above implies that under the assumption from Eq.~\eqref{eq:lambda_app} for sufficiently small $\epsilon$, the map $\e^{\L t'}$ is close to a unitary channel. This presents the $\epsilon$-dependent version of the first case in Proposition~\ref{prop:qubit_simp}. Therefore, applying the map two times is also close to a unitary channel and almost preserves the orthogonality of the evolved states of $\proj0$ and $\proj1$. More precisely, for any time $t$ in $[0,t_f]$,
    \begin{align}
     \!\!\!&D\left(\rho_0(t),\rho_1(t)\right)\geq D\left(\e^{\L\frac{ t}{2}}(\psi_0(t/2)),\e^{\L\frac{ t}{2}}(\psi_1(t/2))\right) \nonumber
        \\ \nonumber
        &\quad-D\left(\rho_0(t),\e^{\L\frac{ t}{2}}(\psi_0(t/2))\right)-D\left(\rho_1(t),\e^{\L\frac{ t}{2}}(\psi_1(t/2))\right)
         \\ \nonumber
         &\geq 1-h(\epsilon)
        -D\left(\rho_0(t/2),\psi_0(t/2)\right)-D\left(\rho_1(t/2),\psi_1(t/2)\right)
        \\
        &\geq1-\left(h(\epsilon)+\epsilon^{0.9}+2\sqrt{2}{\epsilon^{\frac{1}{4}}}\right)=:1-h'(\epsilon),
    \end{align}
    where we used the triangle inequality, and then applied Eq.~\eqref{eq:ortho-preserving} and the data processing inequality to get the second inequality. Finally, the last inequality was obtained using the fact that $\psi_0(t/2)$ is the eigenstate of $\rho_0(t/2)$ corresponding to its largest eigenvalue (recall Eq.~\eqref{eq:lambda_app}), and applying the upper bound from Eq.~\eqref{eq:fidelity-distance_app}, while noting that by Uhlmann's theorem on fidelity, it holds that $ F\left(\rho_1(t/2),\psi_1(t/2)\right)\geq|s_{10}|^2$.
    Above equation therefore  enforces
    \begin{align}\label{eq: first}
       \!\!\!D\left(\rho_1(t_f),\proj{1}\right)&\geq D\left(\rho_1(t_f),\rho_0(t_f)\right) -D\left(\rho_0(t_f),\proj1\right)\nonumber\\
       \!\!\!&\geq \!1-\!h'(\epsilon)\!-\!\sqrt{\epsilon(2-\epsilon)}=:1\!-\! f(\epsilon),\!
    \end{align}
    with $f$ being defined in Eq.~\eqref{subeq:f}. Here, we also employed the following
    \begin{equation}
        \label{eq:rho_distance_to_one}
    D\left(\rho_0(t_f),\proj1\right)\leq\sqrt{\epsilon(2-\epsilon)},
    \end{equation}
    which is a result of Eq.~\eqref{eq:e-neighbour} and the upper bound from Eq.~\eqref{eq:fidelity-distance_app}. Finally, we note that Eq.~\eqref{eq: first} gives a non-trivial bound for any $\epsilon\leq 10^{-6}$ and is saturated only when $\epsilon=0$, which completes the proof of Eq.~\eqref{item:uniatry_app}.

   \textbf{Almost completely contractive dynamics.} For the second part, where we do not use the assumption on purity from Eq.~\eqref{eq:lambda_app}, there exists $t_\star<t_f$ such that
    \begin{equation}\label{eq: eignenvalue of rank two_app}
        \frac12\leq\lambda_{t_\star}<1-\epsilon^{0.9}.
    \end{equation}
    Define $\tilde{t}:=t_f-t_\star$, as well as
    \begin{equation}
    \!\!\! F_0:=\bra1\e^{\tilde{t}\cal L}\left({\psi_0(t_\star)}\right)\ket1,\quad F_1:=\bra1\e^{\tilde{t}\cal L}\left({\psi_1(t_\star)}\right)\ket1,
    \end{equation}
    so that the fidelity $\mathcal{F}$ between $\rho_0(t_f)$ and $\ketbra{1}{1}$ reads:
    \begin{equation}
        \mathcal{F}:=F\left(\rho_0(t_f),\proj1\right)=\lambda_{t_\star}F_0+(1-\lambda_{t_\star})F_1.
    \end{equation}
    Due to the bounds on the dominant eigenvalue in Eq.~\eqref{eq: eignenvalue of rank two_app},
    if $F_0\leq 1-2\epsilon(2-\epsilon)$ or $F_1\leq 1-\epsilon^{0.1} (2-\epsilon)$, then $\mathcal{F}\leq(1-\epsilon)^2$. However, by the upper bound of trace distance based on fidelity from Eq.~\eqref{eq:fidelity-distance_app}, one gets
    \begin{align}
       \!\!\! \F=1\!-\!\bra{0}\rho_0(t_f)\ket{0}\geq D\left(\rho_0(t_f),\proj0\right)^2\geq(1-\epsilon)^2.
    \end{align}
    Therefore, we should have both $F_0\geq  1-2\epsilon(2-\epsilon)$
    and $F_1\geq 1-\epsilon^{0.1} (2-\epsilon)$ to achieve the above bound.
    With the restriction that $\epsilon$ is less than $10^{-6}$, none of these bounds are trivial. For small enough $\epsilon$, this means that the evolution almost sends both $\ket{\psi_0(t_\star)}$ and $\ket{\psi_1(t_\star)}$ to the same pure state. More precisely,
    \begin{align}
       &1-(2-{\epsilon})(\epsilon+{\epsilon^{0.1}}/{2})\nonumber\\
       &\leq\frac12
        \left(F[\e^{\cal L\tilde{t}}\left({\psi_0(t_\star)}\right),\proj1]
        +F[\e^{\cal L\tilde{t}}\left({\psi_1(t_\star)}\right),\proj1]\right)\nonumber\\
        &=
        F\left(\e^{\tilde t\cal L}\left(\frac{\proj{\psi_0(t_\star)}+\proj{\psi_1(t_\star)}}{2}\right),\proj{1}\right)
       \nonumber \\&=
        F(\e^{\tilde t\cal L}\left(\frac{\proj{\xi}+\proj{\xi^\perp}}{2}\right),\proj{1})
       \nonumber \\&=
        \frac12 \left(F(\e^{\tilde t\cal L}(\proj{\xi}),\proj{1})+
        F(\e^{\tilde t\cal L}(\proj{\xi^\perp}),\proj{1})\right),
    \end{align}
    where $\ket{\xi}$ and $\ket{\xi^\perp}$ are any two orthogonal states. The above inequality, together with the facts that fidelity is less than unity and the dynamics is Markovian, implies that for all $t\geq\tilde t$
    and all $\ket{\xi}$ we have
    \begin{eqnarray}
    \label{eq:epsilon_m}
        F(\e^{t\cal L}(\proj{\xi}),\proj{1})\geq 1-(2-{\epsilon})(2\epsilon+\epsilon^{0.1}),
    \end{eqnarray}
    which means that the entire space almost collapses to the neighbourhood of $\proj1$ at $t_f\geq\tilde{t}$. This gives the $\epsilon$-dependent version of the second case in Proposition~\ref{prop:qubit_simp}.
    Applying the upper bound of trace distance based on fidelity in Eq.~\eqref{eq:fidelity-distance_app}, one gets
    \begin{align}
      \!\!\!  D(\e^{\cal L t_f}(\proj{\xi}),\proj1)&\leq \sqrt{(2-{\epsilon})(2\epsilon+\epsilon^{0.1})}=g(\epsilon),
    \end{align}
    which completes the proof.
\end{proof}

    We can now apply the above result to prove Theorem~\ref{thm:qubit}.
    \begin{proof}[Proof of Theorem~\ref{thm:qubit}]
         Take any $0<\epsilon\leq10^{-6}$ and a stochastic matrix $T$ given by Eq.~\eqref{eq:2x2_stochastic} with $a<\epsilon$ and $f(\epsilon)\left(2-f(\epsilon)\right)<b<1-g(\epsilon)$. For such a $T$ to be quantum-embeddable, there has to exist a dynamical map $\e^{\L t_f}$ such that
    \begin{subequations}
        \begin{align}
          \bra{0}\e^{\L t_f}(\proj0)\ket0&\leq\epsilon, \label{subeq:fidelity01}\\
             f(\epsilon)\left(2-f(\epsilon)\right)\leq\bra{1}\e^{\L t_f}(\proj1)\ket1&\leq 1-g(\epsilon)\label{subeq:fidelity11},
        \end{align}
    \end{subequations}
     so that the resulting classical action lies in the immediate neighbourhood of~$T$. Otherwise, one can always find \mbox{$\delta>0$} such that, for any Lindbladian $\L$ and time $t_f$, Eq.~\eqref{eq:Embeddability_definition} is violated. To see that, we apply the lower bound on trace distance given in Eq.~\eqref{eq:fidelity-distance_app} to Eq.~\eqref{subeq:fidelity01} and get
    \begin{equation*}
      D\left(\e^{\L t_f}(\proj0),\proj0\right)\geq1-\epsilon.
    \end{equation*}
    Thus, Theorem~\ref{thm:Qubit-General_app}, along with Eq.~\eqref{eq:fidelity-distance_app}, implies that for any $0<\epsilon\leq10^{-6}$ one of the following has to hold
    \begin{subequations}
        \begin{align}
            F\left(\e^{\L t_f}(\proj1),\proj1\right)&< f(\epsilon)\left(2-f(\epsilon)\right),\label{eq:fidfin1}\\
           F\left(\e^{\L t_f}(\proj1),\proj1\right)&>1-g(\epsilon).\label{eq:fidfin2}
        \end{align}
    \end{subequations}
    Note that in the last equation above, we directly applied Eq.~\eqref{eq:epsilon_m}. We conclude that Eqs.~\eqref{eq:fidfin1}-\eqref{eq:fidfin2} contradict Eq.~\eqref{subeq:fidelity11}, which completes the proof.

    \end{proof}


\subsection{Proof of Theorem~\ref{thm:main}}
\label{sec:qudit_nonembeddable}

We now proceed to proving that for systems of arbitrary dimension $d$ there exist stochastic matrices that are not quantum-embeddable. This will be achieved by Proposition~\ref{thm:orthogonality_evolution} followed by the proof of Theorem~\ref{thm:main}.
\begin{prop}
\label{thm:orthogonality_evolution}
    Let two states $\rho_1$ and $\rho_2$ evolve under Markovian quantum dynamics generated by $\L$, such that  at some time $t_f > 0$ we have
    \begin{equation}
    \label{eq:two_orthogonal_states}
        D\left(\e^{\L t_f}(\rho_1),\e^{\L t_f}(\rho_2)\right)\geq1-\epsilon.
    \end{equation}
    Then, for any $t\geq0$, the following inequality for the Hilbert-Schmidt inner product holds
    \begin{equation}
    \label{eq:inner_product_evolution}
\!\!\!        \Tr\left[\e^{\L (t_f+t)}(\rho_1)\e^{\L(t_f+t)}(\rho_2)\right]\leq H(d)^{\lceil\frac{(d^{4}-1)t}{t_f}\rceil}\epsilon(2-\epsilon),\!
    \end{equation}
    where
    \begin{equation}\label{eq:H(d)}
	    H(d) \coloneqq \frac{(d^4+1)!}{2} d^{d^4+4}.
    \end{equation}
\end{prop}
The proof of the above proposition can be found in Appendix~\ref{app:orthogonality}. Note that one can always take $\epsilon$ small enough so that Eq.~\eqref{eq:inner_product_evolution} gives a non-trivial bound. We employ Proposition~\ref{thm:orthogonality_evolution} here to show, by contradiction, that a stochastic matrix $T$ satisfying the conditions stated in Theorem~\ref{thm:main} is not quantum-embeddable. More precisely, we will prove that one can always find $\delta>0$ such that  Eq.~\eqref{eq:Embeddability_definition} does not hold for such $T$.

\begin{proof}[Proof of Theorem~\ref{thm:main}]
    Assume that a matrix $T$, satisfying the requirements stated in Theorem~\ref{thm:main}, is quantum-embeddable. Being an invariant permutation on $\I_0$ means that for any $\delta>0$ there exist a Lindbladian $\L$ and time $t_f$ such that for any $i_0\in\I_0$ one can find (not necessarily distinct) indices $j_0,k_0\in\I_0$ satisfying
\begin{subequations}\label{eq:subspace_permutation}
    \begin{align}
       \bra{i_0}\e^{\L t_f}(\proj{j_0})\ket{i_0}\geq1-\delta,\label{subeq:j_0 to i_0}\\
       \bra{j_0}\e^{\L t_f}(\proj{k_0})\ket{j_0}\geq1-\delta.\label{subeq:k_0 to j_0}
    \end{align}
\end{subequations}
Moreover, there has to exist an index $i_0\in\I_0$ such that
\begin{subequations}\label{eq:subspace_collapse}
    \begin{align}
       \forall i_1\in\I_1&:\bra{i_0}\e^{\L t_f}(\proj{i_1})\ket{i_0}\geq1-\delta,\label{subeq:subspace_collapse}\\
       \exists i\in\left(\I_0\cup\I_1\right)^c&: \!\sum_{i_1\in\I_1}\!\!\bra{i_1}\e^{\L t_f}(\proj i)\ket{i_1}\geq1-m\delta,\label{subeq:subspace_collapse2}
    \end{align}
\end{subequations}
where $m=|\I_1|$ is the cardinality of $\I_1$.
Using the notation introduced in Eq.~\eqref{eq:simple_notation}, Eqs.~\eqref{subeq:k_0 to j_0}-\eqref{subeq:subspace_collapse2} give
\begin{subequations}\label{eq:i0_and_k0_pi}
    \begin{align}
    \label{eq:i0_and_k1}
       \Tr\left(\Pi \rho_{k_0}(t_f)\Pi\right)&\geq1-\delta,\\
        \Tr\left(\Pi_{\I_1} \rho_i(t_f)\Pi_{\I_1}\right)&\geq1-m\delta,
        \label{eq:i0_and_k2}
    \end{align}
\end{subequations}
where $\Pi = \proj{j_0}$ and $\Pi_{\I_1}$ is the projector onto the subspace $\mathbb{V}_{\I_1}=\mathrm{Span}\{\ket{i_1}\}_{i_1\in\I_1}$. Now, note that, for small enough $\delta$, while the states $\rho_{k_0}(t_f)$ and $\rho_{i}(t_f)$ are almost orthogonal, if we apply the map $\e^{\L t_f}$ to these states, obtaining $\rho_{k_0}(2t_f)$ and $\rho_{i}(2t_f)$, then the resulting states both will be very close to the state $\proj{i_0}$, and therefore very close to each other. The reason is that the state $\proj{j_0}$, as well as the entire subspace $\mathbb{V}_{\I_1}$, collapses to the close neighbourhood of $\proj{i_0}$ because of Eqs.~\eqref{subeq:j_0 to i_0} and \eqref{subeq:subspace_collapse} for small enough $\delta$.

However, introducing $\Pi^\perp= \iden - \Pi $, for any $t\leq t_f$:
\begin{align}
    \nonumber&D\left(\rho_{k_0}(t),\rho_i(t)\right)\geq D\left(\rho_{k_0}(t_f),\rho_i(t_f)\right)\geq\\ \nonumber
    &D\!\left(\Pi\rho_{k_0}(t_f)\Pi\!+\!\Pi^\perp\rho_{k_0}(t_f)\Pi^\perp\!,\Pi\rho_{i}(t_f)\Pi\!+\!\Pi^\perp\rho_{i}(t_f)\Pi^\perp\right)\!=\\
    \nonumber
    &D\!\left(\Pi\rho_{k_0}\!(t_f)\Pi,\Pi\rho_{i}(t_f)\Pi\right)\!+\!D\!\left(\Pi^\perp\rho_{k_0}\!(t_f)\Pi^\perp,\Pi^\perp\rho_{i}(t_f)\Pi^\perp\!\right)\\
    &\geq1-(m+1)\delta,
\end{align}
where we first used the data processing inequality twice, and then we employed Eq.~\eqref{eq:i0_and_k0_pi}. The above, due to Proposition~\ref{thm:orthogonality_evolution}, implies
\begin{equation}
    \!\!\! \Tr\left(\rho_{k_0}(2t_f)\rho_i(2t_f)\right)\!\leq\! H(d)^{d^{4}\!-\! 1}(m\!+\! 1)\delta(2\!-\!(m\!+\! 1)\delta).\!
\end{equation}
This upper bound means that the smaller the $\delta$ is, the farther the states $\rho_{i}(2t_f)$ and $\rho_{k_0}(2t_f)$ are. This contradicts the discussion following Eqs.~\eqref{eq:i0_and_k1}-\eqref{eq:i0_and_k2} and completes the proof of Theorem~\ref{thm:main}.
\end{proof}


\subsection{Proof of Theorem~\ref{thm:embeddable}}
\label{sec:qudit_embeddable}

We now provide  a constructive proof of Theorem~\ref{thm:embeddable}. Without sacrificing generality, assume that $T$ has the form given in Eq.~\eqref{eq:block_form_matrix}  and $R$ of dimension $d'$ is a quantum-embeddable stochastic matrix acting on the levels $\{1,\dots,d'\}$.  Also, for any $j>d'$, let the column $j$ of $T$ be a copy of its column $i_j\leq d'$. Note that this notation allows for two distinct $j,k>d'$ to have $i_j=i_k$, meaning that the columns $j$ and $k$ are both the same copy. In what follows, we show that such a matrix $T$ is quantum-embeddable, i.e., for any $\delta$ there exist a Lindbladian $\L$ and time $t_f$ such that Eq.~\eqref{eq:Embeddability_definition} holds.

Denote by $\v_{d'}$ the $d'$-dimensional subspace spanned by $\{\ket{i}\}_{i=1}^{d'}$, by $\v_{d'}^\perp=\mathrm{Span}\{\ket{i}\}_{i=d'+1}^{d}$ its orthogonal subspace, and by $\Pi_{\v_{d'}}$ and $\Pi_{\v_{d'}^\perp}$ the projectors onto these subspaces, respectively. Since $R$ is assumed to be quantum-embeddable, there has to exist a Lindblad generator $\L_R$ such that the classical action of $\e^{\L_R t_f}$ is arbitrarily close to $R$ for some $t_f$. Such a dynamical map can be chosen to have a trivial action on operators acting on $\v_{d'}^\perp$, i.e., $\L_R(X)=0$ if $X=\Pi_{\v_{d'}^\perp}X\ \Pi_{\v_{d'}^\perp}$. Next, consider a Lindbladian
\begin{equation} \label{eq:complementary_lindbladian}
    \L_S(\cdot)=\sum_{j>d'}\ketbra{i_j}{j}(\cdot)\ketbra{j}{i_j}-\frac{1}{2}\{\Pi_{\v_{d'}^\perp},\cdot\}.
\end{equation}
The Lindbladian $\L_S$ generates a completely dissipative dynamics on the subspace $\v_{d'}^\perp$ that eventually sends each state $\ket{j}$ to $\ket{i_j}$. Trivially, it holds that $\L_S(Y)=0$ if $Y=\Pi_{\v_{d'}}Y\ \Pi_{\v_{d'}}$ acts on $\v_{d'}$.

Therefore, one infers that the Lindbladian \mbox{$\L=\L_R+\gamma\L_S$}, for sufficiently strong coupling $\gamma\gg1$, sends the population of the level $\proj j$ to $\proj{i_j}$ with arbitrary precision in arbitrarily short time. It is because, in the limit $\gamma\rightarrow\infty$, this transformation happens exactly right at the beginning of the evolution. Henceforth, $\gamma$ can be set such that, for any demanded precision $1-\epsilon$, there exists an arbitrarily short time  $t_\star\ll1$ such that
\begin{equation}
    \e^{\L t_\star}(\proj j)= (1-\epsilon)\proj{i_j} +\epsilon\xi_j(t_\star),
\end{equation}
where $\xi_j(t_\star)$ is an arbitrary state. Thus, for the remaining time $\tilde{t}=t_f-t_\star\approx t_f$ the level $\proj{j}$ undergoes approximately the same evolution as the state $\proj{i_j}$. This means that at $t_f$ they are arbitrarily close, which completes the proof of Theorem~\ref{thm:embeddable}.


\subsection{Proof of Corollary~\ref{cor:extreme_points}}
\label{sec:extreme_points}

We will now argue why Corollary~\ref{cor:extreme_points} is a straightforward consequence of Theorems~\ref{thm:main} and \ref{thm:embeddable}. This will be achieved through the following lemma.
\begin{lem}
\label{lem:extrem-permutations}
    Let $T$ be an extreme stochastic matrix. Then, one can always find a set $\I_0$ of indices such that $T$ invariantly permutes $\I_0$ and, when $\I_0^c$ is non-empty, there exists $i\in\I_0^c$ which is sent by $T$ to $\I_0$.
\end{lem}
\begin{proof}
  Since $T$ is an extreme stochastic matrix, for any~$i_0$ one can define the set $S_{i_0}$ as the  largest possible set of distinct indices obtained  by acting with $T$ sequentially on $i_0$, i.e.,  $S_{i_0}=\{i_0,i_1,\dots,i_{n_{i_0}}\}$ where $T^m_{i_m,i_0}=1$ for any $m\in\{0,1,\dots,n_{i_0}\}$, while $T^{n_{i_0}+1}_{j,{i_0}}=T_{j,i_{n_{i_0}}}=\delta_{j,i_k}$ for some $k\leq n_{i_0}$. There are two possibilities. If for any $i_0$ we get $i_k=i_0$, i.e., $T_{i_0,i_{n_{i_0}}}=1$, then $T$ is a permutation and proves the lemma. On the other hand, if there exists an $i_0$ for which $i_k\neq i_0$, then $T$ is a permutation on $\I_0=S_{i_k}=\{i_k,\dots,i_{n_{i_0}}\}$ and $i_{k-1}$ is mapped to $S_{i_{k}}$, which completes the proof.
\end{proof}
\begin{proof}[Proof of Corollary~\ref{cor:extreme_points}]
    To prove Corollary~\ref{cor:extreme_points}, we note that since $T$ is  assumed to be an extreme stochastic matrix, through Lemma~\ref{lem:extrem-permutations}, there always exists $\I_0$  which $T$ invariantly permutes. If $T$ is a permutation, then it is quantum-embeddable. If it is not a permutation, then we can use the notation introduced in the proof of Lemma~\ref{lem:extrem-permutations} to show that there exist indices $i_0$ such that $i_k\neq i_0$. For these indices, there are only two possibilities. Either for all $i_0$ with $i_k\neq i_0$ we get $k=1$, which is the structure posed by Theorem~\ref{thm:embeddable} and gives a quantum-embeddable map as a result. Or, otherwise, there exists an index $i_0$ such that $k\geq2$, implying that \mbox{$T_{i_{k-1},i_{k-2}}=T_{i_{k},i_{k-1}}=1$}, where $i_k\in\I_0$. This is the structure given by Theorem~\ref{thm:main} and yields a non-quantum-embeddable map.
\end{proof}


\section{Conclusions and outlook}
\label{sec:conclusions}

In this work, we investigated the set of quantum-embeddable stochastic matrices, i.e., classical state transition maps that arise from time-homogeneous quantum Markov dynamics. For the dimension $d=2$, in Theorem~\ref{thm:qubit}, we provided an analytical description of a curve that outer approximates the set $\Q_2$ of $2 \times 2$ quantum-embeddable stochastic matrices (recall Fig.~\ref{fig:qubit_regions}). In particular, our result implies that the set of not quantum-embeddable matrices $\Q_2^c$ has non-zero volume within $\T_2$. For higher dimensions $d>2$, we derived non-trivial inner and outer approximations on the set $\Q_d$. To achieve this, we bounded a ratio $\epsilon$ at which time-homogeneous memoryless quantum channels $\epsilon$-preserve orthogonality of input states (Proposition~\ref{thm:orthogonality_evolution}). As a consequence, in Theorem~\ref{thm:main}, we were able to characterise some elements of $\Q_d^c$, thus outer approximating $\Q_d$. Moreover, by mixing dissipative dynamics and unitary evolution, we constructed a new class of quantum-embeddable matrices (Theorem~\ref{thm:embeddable}) that goes beyond classically-embeddable matrices $\C_d$ and unistochastic matrices $\U_d$, and thus provides a new inner approximation on $\Q_d$. Finally, by combining the results from Theorem~\ref{thm:main} and Theorem~\ref{thm:embeddable}, we comprehensively characterised all extreme stochastic matrices that are quantum-embeddable (see Corollary~\ref{cor:extreme_points}).

Concerning our technical results, there is still plenty of room for improvement. First, the numerical investigation provided in Appendix~\ref{app:numerics} shows that the boundary of $\Q_2$ differs from the one derived in Theorem~\ref{thm:qubit} by a few orders of magnitude. One way to improve the theoretical bound could be to prove that the numerically revealed $4$-dimensional family of Lindbladian operators indeed generates the boundary of the set $\Q_2$. Second, it would be interesting to estimate the volume of $\Q_d^c$ for arbitrary $d$. In Theorem~\ref{thm:main}, we characterised a particular type of stochastic matrices belonging to $\Q_d^c$ and so, remembering that $\Q_d$ is closed, one may try to find non-zero balls of non-quantum-embeddable matrices around these matrices, and provide lower bound on the volume of $\Q_d^c$ by estimating the radius of such balls. Finally, one might also try to devise a systematic approach to constructing new families of quantum-embeddable stochastic matrices, which would yield a subset on the set $\Q_d$.

On the other hand, an alternative avenue is to apply a different map to encode a stochastic matrix in a quantum channel rather than Eq.~\eqref{eq:E to T}. This can be done according to the practical intentions and available experimental settings. For example, if preparation and measurement in arbitrary basis sets are possible, a stochastic matrix $T$ can be defined by $   T_{ij}=\bra{x_i}\E(\proj{y_j})\ket{x_i}$,
where $\{\ket{x_i}\}$ and $\{\ket{y_i}\}$ are two sets of orthonormal basis. Now, one can pose the quantum embeddability question for the above definition. This is equivalent to asking for a given $T$ whether there exist a time-homogenous Markovian quantum channel $\E$ and unitary channels $\U$ and $\V$ such that $T_{ij}=\bra{i}\ \U\circ\E\circ\V\left(\proj{j}\right)\ket{i}$. The set of all such transitions trivially contains $\Q_d$ as a subset.
In the $2\times2$ case, it is easy to check by choosing $\{\ket{x_i}\}$ and $\{\ket{y_i}\}$ as permutations of the set $\{\ket0,\ket1\}$, the entire set of stochastic matrices is recovered. However, $\U\circ\E\circ\V$, for a Markovian $\E$, is a concatenation of three Markovian channels, and therefore, it is time-inhomogenous Markovian itself. However, such a concatenation, being very specific, does not necessarily give the entire set of time-inhomogenous Markovian channels. Thus, the set of quantum embeddable matrices, even with the freedom in preparation and measurement basis, is just a subset of the embeddable maps given by the time-inhomogenous dynamics. The latter is a proper subset of stochastic matrices in dimensions $d>2$~\cite{korzekwa2021quantum}.

An alternative encoding process is to apply a projective measurement with $n$ outcomes on the outputs of the channel acting on one of $n$ states of dimension $d$, where $d>n$. That is to say, $T_{ij}=\Tr\left(\Pi_i\E(\rho_j)\right)$ where $i,j\in\{1,\dots,n\}$, $\{\Pi_i\}$ is a set of $n$ projective measurements of size $d$ and $\{\rho_j\}$ is a set of $n$ input states of size $d$. This definition is expected to significantly enlarge the set of embeddable maps. The reason is that the additional degrees of freedom can indeed play the role of the memory. It is easy to show that in the extreme case, the entire set of $n$-dimensional stochastic matrices is recovered by quantum unitary dynamics of dimension $d=n^2$ and projective measurements of rank $n$. To see that, one can take the set of input states as $\{\rho_j\}=\{\proj{jn}\}$ and the rank-$n$ projector as $\Pi_i=\sum_{m=0}^{n-1}\proj{i+mn}$ for $i,j\in\{0,\dots,n-1\}$. Then, for $k\in\{0,\dots,n-1\}$ the unitary dynamics given by $U=\bigoplus u_k$ with a proper choice of unitary operators $u_k$ acting on subspaces $V_k=\mathrm{Span}\{\ket{kn},\ket{kn+1},\cdots,\ket{kn+n-1}\}$ generates an arbitrary $n$-dimensional stochastic matrix $T$.

On a more conceptual level, our research reveals the inherent advantages offered by quantum dynamics over the classical dynamics in the context of generating stochastic processes without using memory.
In a related, yet different framework, it has been shown that quantum mechanics offers both entropy~\cite{verdral-complexity-12} and dimensional~\cite{verdral-complexity-14,ghafari2019dimensional} memory advantages for simulating the unifilar representation of stochastic processes. In practice, simulating such processes requires some bipartite quantum interaction to be applied over an $s$-dimensional ancillary memory and an $r$-dimensional system to generate a single output of the process, where $r$ represents the number of the outputs of the process and $s$ is the number of causal states (i.e., the minimum effective number of inputs required to generate the process). To generate the next step of the process, one then needs to keep the memory intact and add another input state to the interaction (for clearer visualization, see Figs. 1  of Refs.~\cite{ghafari2019dimensional} and~\cite{binder2018practical}).

Here, the results cover a special stochastic process, namely, a $d$-dimensional biased coin. For a generic case of this process, the number of outputs $r$ and causal states $s$ coincide with $d$. If a stochastic matrix related to such a process is classically-embeddable, then it
can be realised by a $d$-dimensional memoryless interaction, implying that there is no necessity to retain the state of the memory. To obtain the next output of the process, the current system state is then reused. In this paper, we showed that if a quantum evolution is allowed instead, one can realise a strictly larger set by the $d$-dimensional memoryless interactions. Therefore, the advantage we propose here is in terms of a realisation of the process. Moreover, as discussed above, if we let $d^2$-dimensional interaction, all such processes can be realised by unitary evolutions that is in agreement with the known results of~\cite{binder2018practical}.

It is clear that the observed advantages in this approach arise from the fact that quantum systems can evolve coherently and thus experience interference effects, and so it would be interesting to quantify the impact of quantum coherence on memory improvements (e.g., how much coherence is necessary to simulate one additional memory state). One idea for that would be to investigate how these improvements behave under decohering noise, e.g., if on top of the Markovianity condition, we require quantum channels generating our stochastic transitions to have a level of noise above some fixed threshold.


\subsection*{Acknowledgements}

FS and KK would like to thank Adam Burchardt, Frederik vom Ende, and Karol \.{Z}yczkowski for fruitful discussions and comments. KK, FS and \LL{}P acknowledge financial support by the Foundation for Polish Science through TEAM-NET project (contract no. POIR.04.04.00-00-17C1/18-00). RK acknowledges financial support by the National Science Centre, Poland, under the contract
number 2021/03/Y/ST2/00193 within the QuantERA II Programme that has
received funding from the European Union’s Horizon 2020 research and
innovation programme under Grant Agreement No 101017733 and acknowledges additionally support by the Foundation for Polish Science through TEAM-NET project (contract no. POIR.04.04.00-00-17C1/18-00). CTC acknowledges support from the Swiss National Science Foundation through the Sinergia grant CRSII5-186364, and for the NCCRs QSIT and SwissMAP. CTC also thanks Jagiellonian University for their hospitality during the realisation of this project.

\appendix


\section{Numerical optimisation over Markovian qubit channels}
\label{app:numerics}

We obtained the curve defining the region $\Q_2$ in Fig.~\ref{fig:qubit_regions} by verifying if
stochastic matrices
\begin{equation}
    T = \begin{pmatrix}
        a&1-b\\
        1-a&b
\end{pmatrix}
\end{equation}
defined for $(a,b) \in [0,1] \times [0,1]$ satisfy the condition in Eq.~\eqref{eq:Embeddability_definition} for some Markovian channel $\mathcal{E}$. The fulfillment of this condition has been checked for $\delta = 10^{-4}$. For each tested point $(a,b)$, we minimised the expression $\max_{i,j}\big|T_{ij}-\matrixel{i}{e^{\L t}(\ketbra{j}{j})}{i}\big|$ over all Lindblad generators $\L$ of the form shown in Eq.~\eqref{eq:lindbladian} and $t > 0$. Without loss of the generality, we parameterised $H$ with a single variable $h \in \R$,
\begin{equation}
    H = \begin{pmatrix}
        \cos h & \sin h\\
        \sin h & -\cos h
\end{pmatrix}.
\end{equation}
The map $\Phi$ was chosen by its Choi-Jamio{\l}kowski isomorphism in the form $J_\Phi = GG^\dagger$, where $G$ is a $4 \times 4$ complex matrix, which introduces additional 32 real parameters.
We carried out the optimisation using the Nelder-Mead optimisation method with random initialisation.
 The numerical investigation revealed  that the boundary curve in Fig.~\ref{fig:qubit_regions} (b) can be achieved for a particular choice of $H$ and $\Phi$. We observed that it is sufficient to consider the Hermitian operation given by the Pauli $X$ operator, $H = \sigma_x$. Moreover, the map $\Phi$ may be defined by a single Kraus operator $A$, $\Phi(\rho) = A\rho A^\dagger$, such that $A = \sqrt{\gamma} \ketbra{\psi_o}{\psi_i}$, where $\ket{\psi_o} = (\cos(\alpha), i \sin(\alpha))$ and $\ket{\psi_i} = (\cos(\beta), i\sin(\beta))$ for $\alpha, \beta \in \R$ and $\gamma \in \R_+$. This simplification reduces the number of optimisation parameters to $\alpha, \beta \in \R$ and $\gamma, t \in \R_+$.

The code to generate the numerical results is available on GitHub~\cite{code}.

\section{Proof of Eq. (\ref{eq:coefficients})}
\label{app:s10}

To prove Eq.~\eqref{eq:coefficients},  we will exploit the entanglement of a $2\times N$ pure bipartite state
$\ket{\Omega}=\sum_{i=0}^1\sum_{j=0}^{N-1}f_{ij}\ket{a_i}\ket{b_j}$, measured by concurrence~\cite{Akhtarshenas2005}:
\begin{equation}\label{eq:concurrence}
    E(\Omega)=2\sqrt{\sum_{j<k}\abs{f_{0j}f_{1k}-f_{1j}f_{0k}}^2}=2\sqrt{\lambda(1-\lambda)},
\end{equation}
where $\lambda$ is an eigenvalue of the marginal state $\Tr_2(\Omega)$.
Moreover, we will make use of the following technical lemma.
\begin{lem}\label{lem:half}
    Let a qubit dynamical map be $\epsilon$-purity-preserving on a state $\ket0$ for some time $t_f$, with $\psi_t$ being the pure state in the neighbourhood of $\rho_0(t)$ at each moment of time. Also, consider that this map sends the quantum state $\ket0$ to the $\epsilon'$-neighbourhood of $\ket1$ at $t_f$, i.e.,
    \begin{equation}
        D\left(\rho_{0}(t_f),\proj1\right)=D\left(\e^{\L t_f}(\proj 0),\proj 1\right)\leq \epsilon'.
    \end{equation}
    Writing $\ket{\psi _t}=c_t\ket0+c^\perp_t\ket1$, then $|c^\perp_{t_f/2}|$ belongs to the interval $[\frac{1}{\sqrt{2}}-(2\epsilon+\epsilon'),1]$ if $2\epsilon+\epsilon'\leq1/\sqrt{2}$.
\end{lem}
\begin{proof}
   To prove, we note that at $t=t_f/2$
    \begin{align}
       \nonumber\!\!\!\!\!|c_{t_f/2}|=&\ D\left(\proj1,\psi_{t_f/2}\right)\leq D\left(\proj1,\rho_{0}(t_f)\right)\\ \nonumber
       &+D(\rho_{0}(t_f),\rho_{0}({t_f}/{2}))+D(\rho_{0}({t_f}/{2}),\psi_{t_f/2})\\ \nonumber
       \leq&\ \epsilon'+D(\rho_{0}({t_f}/{2}),\proj0)+D(\rho_{0}({t_f}/{2}),\psi_{t_f/2})\\ \nonumber
       \leq&\ \epsilon'+D\left(\psi_{t_f/2},\proj0\right)+2D(\rho_{0}({t_f}/{2}),\psi_{t_f/2})\\
       \leq&\ |c^\perp_{t_f/2}|+2\epsilon+\epsilon',
    \end{align}
    where the first and the third inequalities are due to the triangle inequality, and  the second is a result of the data processing inequality. Applying the above, as long as $2\epsilon+\epsilon'\leq1$, we straightforwardly get
    \begin{equation}\label{eq:MoreThanHalf}
        \abs{c^\perp_{t_f/2}}\geq \frac{1}{\sqrt{2}}-(2\epsilon+\epsilon').
    \end{equation}
    However, this is a non-trivial bound for $2\epsilon+\epsilon'\leq\frac{1}{\sqrt{2}}$.
\end{proof}

Having  Eqs. \eqref{subeq:evolved0}-\eqref{subeq:evolved0+1} in mind, to prove Eqs.~\eqref{eq:s00_main}-\eqref{eq:s10bound_main}, we start by perceiving that due to the orthogonality of $\ket{\Psi_0(t')}$ and $\ket{\Psi_1(t')}$ one has
     \begin{equation}
        \!\!\!\!\!|s_{00}(t')|=\sqrt{\frac{1\!-\!\lambda_{t'}}{\lambda_{t'}}}|s_{11}(t')|\leq\sqrt{\frac{1\!-\!\lambda_{t'}}{\lambda_{t'}}(1\!-\!|s_{00}(t')|^2)}.\!
    \end{equation}
    The above, together with Eq.~\eqref{eq:lambda_app}, results in
     \begin{equation}\label{eq:s00}
        |s_{00}(t')|=\sqrt{1-\lambda_{t'}}\leq\sqrt{\epsilon^{0.9}}=:h_0(\epsilon),
    \end{equation}
    and proves Eq.~\eqref{eq:s00_main}. To proceed with the proof, we note that Eq.~\eqref{eq:lambda_app}, along with Lemma~\ref{lem:trajectory}, restricts at any $t,t'\in [0,t_f/2]$ the concurrence of $\ket{\Psi_{\!\psi_0\!(t)}(t')}$ to
     \begin{align}
           E\left(\Psi_{\!\psi_0\!(t)}(t')\right)&=2\sqrt{\lambda_{t,t'}\left(1-\lambda_{t,t'}\right)}\leq 2\sqrt{2\epsilon^{0.9}},
     \end{align}
     where we used $\lambda_{t,t'}$ to denote the largest eigenvalue of~$\rho_{\!\psi_0\!(t)}(t')$. Recall that through Eq.~\eqref{eq:concurrence} we also get
    \begin{equation}
        \!\! E\left(\Psi_{\!\psi_0\!(t)}(t')\right)\!=\!2\sqrt{\sum_{j<k}\abs{f_{0j}f_{1k}\!-\! f_{1j}f_{0k}}^2}\leq 2\sqrt{2\epsilon^{0.9}},
    \end{equation}
    which implies that each term is bounded by
    \begin{align}\label{eq:Gepsilon}
       \!\!\!\! \abs{f_{0j}(t,t')f_{1k}(t,t')-f_{1j}(t,t')f_{0k}(t,t')}\leq\sqrt{2\epsilon^{0.9}},
    \end{align}
    for any $j,k$ such that $j<k$.
    On the other hand, writing $\ket{\psi_0(t)}=c_t\ket0+c^\perp_t\ket1$, we can restate Eq.~\eqref{subeq:evolved0+1} as
    \begin{equation}
        \ket{\Psi_{\!\psi_0\!(t)}(t')}=c_t\ket{\Psi_0(t')}+c^\perp_t\ket{\Psi_1(t')}.
    \end{equation}
    Therefore,
    \begin{subequations}
        \begin{align}
            f_{00}(t,t')&=c_t\sqrt{\lambda_{t'}}+c^\perp_t s_{00}(t'),\\
            f_{11}(t,t')&=c_t\sqrt{1-\lambda_{t'}}+c^\perp_t s_{11}(t'),\\
            f_{jk}(t,t')&=c^\perp_t s_{jk}(t'), \qquad\mathrm{otherwise}.
        \end{align}
    \end{subequations}
     Hereafter, we may drop writing explicitly the dependence on $t'$ through $\lambda_{t'}$ and $s_{jk}(t')$ for brevity. The above gives six inequalities for different choices of $j<k$ due to Eq.~\eqref{eq:Gepsilon}, which particularly include the following three for the case of $j=0$:
    \begin{subequations}\label{eq:g-inequlity}
        \begin{align}
            \!\!\!\!\label{subeq:2}\sqrt{2\epsilon^{0.9}}\geq& \ \abs{c_t^\perp}\abs{(c_t\sqrt{\lambda}+c_t^\perp s_{00})s_{1k}-c_t^\perp s_{10}s_{0k}},\\
           \!\!\!\!\label{subeq:4}\sqrt{2\epsilon^{0.9}}\geq &\ \Big|(c_t\sqrt{\lambda}+c_t^\perp s_{00})(c_t\sqrt{1-\lambda}+c_t^\perp s_{11})\nonumber\\
           &~-(c_t^\perp)^2s_{10}s_{01}\Big|,
        \end{align}
    \end{subequations}
    where $k=2,3$. The above inequalities are valid for any \mbox{$t,t'\in[0,t_f/2]$}.

    On the other hand, it has been shown that the unitary evolution in the Stinespring dilation of a Markovian dynamics is continuous~\cite{DivePRA15Smoothness}. Thus, Lemma~\ref{lem:half} and Eq.~\eqref{eq:rho_distance_to_one}, by virtue of continuity of the evolution, imply that there has to exist $t_\ast\leq t_f/2$ such that for different times in $[0,t_\ast]$, the coefficient $|c^\perp_t|$ can get any value in $[0,a(\epsilon)]$ where
    \begin{equation}
        a(\epsilon)=\frac{1}{\sqrt{2}}-\left(2\epsilon^{0.9}+\sqrt{\epsilon(2-\epsilon)}\right).
    \end{equation}
    This gives a meaningful interval if $\epsilon\leq0.1$, which introduces the first bound on $\epsilon$. Thus, the inequalities in Eqs.~\eqref{subeq:2}-\eqref{subeq:4} hold for any $t'\leq t_f/2$ and for any $|c^\perp_t|\in[0,a(\epsilon)]$.
    Therefore, from Eq.~\eqref{subeq:2} we have the following
    \begin{align}
    \!\!\!  \sqrt{2\epsilon^{0.9}}&\geq \abs{c_t^\perp}\!\cdot\!\abs{(c_t\sqrt{\lambda}+c_t^\perp s_{00})s_{1k}-c_t^\perp s_{10}s_{0k}} \nonumber\\
        &\geq\abs{c^\perp_t}\Big|\!\cdot\! \abs{c^\perp_t(s_{00}s_{1k}- s_{10}s_{0k})}-\sqrt{\lambda}\abs{c_ts_{1k}}\Big|,
    \end{align}
  where for two specific values, $\abs{c^\perp_t}=\frac12$ and $\abs{c^\perp_t}=a(\epsilon)$, we get
    \begin{subequations}
        \begin{align}
             \label{eq:subb1}\sqrt{2\epsilon^{0.9}}&\geq \frac{\sqrt{3}}{4}\abs{s_{1k}}\sqrt{\lambda}-\frac{1}{4}\abs{s_{00}s_{1k}- s_{10}s_{0k}},\\
             \label{eq:subb2}\sqrt{2\epsilon^{0.9}}&\geq a^2(\epsilon)\abs{s_{00}s_{1k}- s_{10}s_{0k}}-\frac{1}{2}\abs{s_{1k}}\sqrt{\lambda}.
        \end{align}
    \end{subequations}
    Assuming that
    \begin{equation}\label{eq:first-bound}
        a^2(\epsilon)>\frac{1}{2\sqrt{3}},
    \end{equation}
    which gives the second  restriction on $\epsilon$,  i.e., \mbox{$\epsilon\leq9.7\times10^{-3}$},  we insert Eq.~\eqref{eq:subb2} into Eq.~\eqref{eq:subb1} and get
    \begin{equation}\label{eq:ConcurrenceTerm2}
        \abs{s_{00}(t')s_{1k}(t')- s_{10}(t')s_{0k}(t')}\leq h_1(\epsilon),
    \end{equation}
    with
    \begin{equation}\label{eq:h2}
        h_1(\epsilon):=\frac{\sqrt{2\epsilon^{0.9}}(1+\frac{2}{\sqrt{3}})}{a^2(\epsilon)-\frac{1}{2\sqrt{3}}}.
    \end{equation}

    This, in turn, bounds $\abs{s_{1k}(t')}$ for $k=2,3$ at any \mbox{$t'\leq t_f/2$} through Eq.~\eqref{eq:subb1} as
     \begin{equation}\label{eq:s1k}
        |s_{1k}(t')|\leq\frac{4\sqrt{2\epsilon^{0.9}}+h_1(\epsilon)}{\sqrt{3(1-\epsilon^{0.9})}}=:h_2(\epsilon).
    \end{equation}
    The above, along with Eqs.~\eqref{eq:s00} and \eqref{eq:ConcurrenceTerm2}, implies
    \begin{equation}\label{eq:h_3}
        |s_{10}(t')s_{0k}(t')|\leq h_1(\epsilon)+h_0(\epsilon)h_2(\epsilon)=:h_3(\epsilon).
    \end{equation}
    This obliges at least one of the following inequalities for $k=2,3$ and $t'\in[0,t_f/2]$:
        \begin{align}
        \!\!\!    |s_{10}(t')|\leq\sqrt{h_3(\epsilon)},
        \quad\mathrm{or}\quad
            |s_{0k}(t')|\leq\sqrt{h_3(\epsilon)}.\label{eq:s0k}
        \end{align}
    Moreover, from Eq.~\eqref{subeq:4} we obtain
    \begin{align}
         \sqrt{2\epsilon^{0.9}}&\geq\abs{c_t^\perp}\big|(c_t\sqrt{\lambda}+c^\perp_ts_{00})s_{11}-c^\perp_t s_{10}s_{01}\big|\nonumber\\ \nonumber
         &-\sqrt{1-\lambda}\abs{c_t(c_t\sqrt{\lambda}+c^\perp_ts_{00})}\\ \nonumber
         &\geq\abs{c^\perp_t}\Big| \abs{c^\perp_t(s_{00}s_{11}- s_{10}s_{01})}-\sqrt{\lambda}\abs{c_ts_{11}}\Big|\\
         &-\sqrt{\epsilon^{0.9}}\left(|c_t|^2+|c_tc_t^\perp|h_0(\epsilon)\right),
    \end{align}
    where in the last inequality we applied Eqs.~\eqref{eq:lambda_app}~and~\eqref{eq:s00}.
    From that, for two special cases of  $\abs{c^\perp_t}=1/2$ and \mbox{$\abs{c^\perp_t}=a(\epsilon)$}, we get
    \begin{subequations}  \label{eq:second}
        \begin{align}
             \label{eq:subb11}\!\!\!\!\!\!\sqrt{2\epsilon^{0.9}}&\geq \frac{\sqrt{3}}{4}\abs{s_{11}}\sqrt{\lambda}-\!\frac{1}{4}\abs{s_{00}s_{11}\!- s_{10}s_{01}}-\!\sqrt{\epsilon^{0.9}},\\
             \label{eq:subb22}\!\!\!\!\!\!\sqrt{2\epsilon^{0.9}}&\geq a^2(\epsilon)\abs{s_{00}s_{11}\!- s_{10}s_{01}}\!-\!\frac{1}{2}\abs{s_{11}}\sqrt{\lambda}\!-\!\sqrt{\epsilon^{0.9}}.
        \end{align}
    \end{subequations}
    Therefore, the same assumption as in Eq.~\eqref{eq:first-bound} leads to
    \begin{equation}\label{eq:ConcurrencTerm4}
        \abs{s_{00}(t')s_{11}(t')- s_{10}(t')s_{01}(t')}\leq h_4(\epsilon),
    \end{equation}
    where
    \begin{equation}\label{h_4}
        h_4(\epsilon):=\frac{(\sqrt{2\epsilon^{0.9}}+\sqrt{\epsilon^{0.9}})(1+\frac{2}{\sqrt{3}})}{a^2(\epsilon)-\frac{1}{2\sqrt{3}}}.
    \end{equation}
    The same approach as before gives from Eqs.~\eqref{eq:s00}, \eqref{eq:second}, \eqref{eq:ConcurrencTerm4} the following:
    \begin{subequations}\label{eq:Second}
    \begin{align}
        |s_{11}(t')|\leq\frac{4(\sqrt{2\epsilon^{0.9}}+\sqrt{\epsilon^{0.9}})+h_4(\epsilon)}{\sqrt{3(1-\epsilon^{0.9})}}=:h_5(\epsilon),\label{eq:s11}\\ \label{eq:h8}
         |s_{10}(t')s_{01}(t')|\leq h_4(\epsilon)+h_0(\epsilon)h_5(\epsilon)=:h_6(\epsilon),
    \end{align}
    \end{subequations}
    which in turn means either
        \begin{align}
            |s_{10}(t')|\leq\sqrt{h_6(\epsilon)},
            \quad\mathrm{or}\quad
            |s_{01}(t')|\leq\sqrt{h_6(\epsilon)}.\label{eq:s10}
        \end{align}

    However, we show that it is impossible to upper bound $|s_{10}(t')|$ by a function of $h_3(\epsilon)$ or $h_6(\epsilon)$ as in Eqs.~\eqref{eq:s0k} and \eqref{eq:s10}. In this order, note that for $i\in\{1,\dots,6\}$ the functions $h_i(\epsilon)$ converge to zero as $\epsilon$ goes to zero and $h_6(\epsilon)\geq h_3(\epsilon)$. Moreover, $s_{ij}(t')$ are continuous with respect to time while at $t'=0$ we have $s_{10}(0)=1$. Consequently, if there exists $t'=t_\ast$ when for the first time $|s_{10}|$ takes the value $\sqrt{h_6}$, then in the best case scenario, according to Eqs.~\eqref{eq:h_3} and \eqref{eq:h8}, $|s_{01}(t_\ast)|=\sqrt{h_6}$ and $|s_{0k}(t_\ast)|=\frac{h_3}{\sqrt{h_6}}$ for $k=2,3$. This, however, means that the moduli of all the coefficients $|s_{ij}(t_\ast)|$ are upper bounded by some functions of $\epsilon$ that monotonically go to zero as $\epsilon$ does (note that $h_3(\epsilon)\leq h_6(\epsilon)$). This, for sufficiently small~$\epsilon$, contradicts the fact that $\ket{\Psi_1(t_\ast)}$ is a normalised state. Thus, it must be that $|s_{10}(t')|\geq\sqrt{h_6(\epsilon)}\geq\sqrt{h_3(\epsilon)}$.
   More precisely, applying Eqs.~\eqref{eq:s00},  \eqref{eq:s1k}, \eqref{eq:h_3}, and \eqref{eq:Second} to the normalisation condition,
   \begin{equation}
       |s_{10}|^2=1-\sum_{ij\neq10}|s_{ij}|^2,
   \end{equation}
   one can verify that for any $\epsilon\leq 10^{-6}$, which is the third and the strongest restriction on $\epsilon$, we have
    \begin{align}\label{eq:s10bound}
    |s_{10}(t')|&\geq\sqrt{\frac{1}{2}\left(A+\sqrt{A^2-4B^2}\right)}\nonumber\\
    &\geq 1-4\sqrt{\epsilon}\geq\sqrt{h_6(\epsilon)},
    \end{align}
    where $A=1-h_0(\epsilon)^2-2h_2(\epsilon)^2-h_5(\epsilon)^2$ and \mbox{$B=2h_3(\epsilon)^2+h_6(\epsilon)^2$}. Thus, $|s_{10}(t')|$ is the dominant coefficient for any $t'\leq t_f/2$ and the equality holds only for $\epsilon=0$, which completes the proof of Eq.~\eqref{eq:s10bound_main}.


\section{Proof of Eqs.~\eqref{eq:ortho-preserving} and \eqref{eq:purity-proof}}
\label{app:orthogonality-purity}

    To prove Eqs.~\eqref{eq:ortho-preserving} and \eqref{eq:purity-proof}, we apply the map $\e^{\L t'}$ on the states $\ket\upsilon=b\ket0+b_\perp\ket1$ and $\ket{\upsilon^\perp}=b^\ast_\perp\ket0-b^\ast\ket1$ through its Stinespring picture. Employing the same notation as in Eqs.~\eqref{subeq:evolved0}-\eqref{subeq:evolved0+1}, for $m=\{\upsilon,\upsilon^\perp\}$ we get
    \begin{align}\label{app-eq:evolved01}
         \ket{\Psi_m(t')}\!:=&\ U(t')\ket{m}\ket{ 0}\\ \nonumber
         =&\ \|\tilde{\chi}_m(t')\|\!\cdot\! \ket{\chi_m(t')}\ket{r_0(t')}\!+\!\|\tilde{\Gamma}_{\!m}(t')\|\!\cdot\! \ket{\Gamma_{\!m} (t')},
    \end{align}
     where the single party state $\ket{\chi_m(t')}=\ket{\tilde{\chi}_m(t')}/\|\tilde{\chi}_m(t')\|$ is the normalised form of
     \begin{subequations}  \label{app-eq:arbit-chi}
        \begin{align}
        \label{app-eq:arbit-chi1} \!\!\!\!&\ket{\tilde{\chi}_\upsilon(t')}\!=\!(b\sqrt{\lambda}\!+\!b_\perp s_{00})\ket{\psi_0(t')}\!+\!b_\perp s_{10}\ket{\psi_1(t')},\\
         \!\!\!\!&\ket{\tilde{\chi}_{\upsilon^\perp}\!(t')}\!=\!(b^\ast_{\!\perp}\sqrt{\lambda}\!-\!b^\ast s_{00})\ket{\psi_0(t')}\!-\!b^\ast s_{10}\ket{\psi_1(t')}.
         \label{app-eq:arbit-chi2}
        \end{align}
     \end{subequations}
     The bipartite state $\ket{\Gamma_{m}(t')}\!=\!\ket{\tilde{\Gamma}_{\!m}(t')}/\|\tilde{\Gamma}_{\!m}(t')\|$ of the system and environment,  for both $m=\{\upsilon,\upsilon^\perp\}$, contains the remaining terms in the expansion of the state $\ket{\Psi_{m}(t')}$ in the basis $\{\ket{\psi_i(t')}\ket{r_j(t')}\}$. Note that, by definition $\bra{\Gamma_{m}(t')}\chi_m(t')\ r_0(t')\rangle=0$, which implies
    \begin{equation}
        \|\tilde{\Gamma}_{\!m}(t')\|^2=1-\|\tilde{\chi}_{m}(t')\|^2\quad\mathrm{for}\quad m\in\{\upsilon,\upsilon^\perp\}.
    \end{equation}
    Using the bounds from Eqs.~\eqref{eq:lambda_app} and \eqref{eq:s00_main}-\eqref{eq:s10bound_main}, one can show that
       \begin{align}\label{app-eq:chi-norm}
           \|\tilde{\chi}_{m}(t')\|^2&\geq 1-(8\sqrt{\epsilon}+\epsilon^{0.45}).
       \end{align}
    Moreover, $\rho_m(t')=\e^{\L t'}(\proj m)$ for  $m=\{\upsilon,\upsilon^\perp\}$ is the marginal of $\proj{\Psi_m(t')}$.

    Now, the proof of both $\epsilon$-purity and $\epsilon$-orthogonality of any $\rho_{\upsilon}$ and $\rho_{\upsilon^\perp}$ is straightforward. Here, we first prove Eq.~\eqref{eq:purity-proof} and then \eqref{eq:ortho-preserving}. For the first, we have the following bound on the largest eigenvalue $\eta$ of $\rho_{m}(t')$:
    \begin{align}\label{app-eq:purity-proof}
        \eta&\geq\bra{\chi_{m}(t')}\rho_{m}(t')\ket{\chi_m(t')}=F\left(\chi_m(t'),\rho_{m}(t')\right)\nonumber\\
        &\geq\big|\bra{\chi_m(t')\ r_0(t')}\Psi_{\!m}(t')\rangle\big|^2=\|\tilde{\chi}_{m}(t')\|^2\nonumber\\
        &\geq1-(8\sqrt{\epsilon}+\epsilon^{0.45}),
    \end{align}
    where the first inequality is based on the fact that the largest eigenvalue of a state is its maximum expectation value with respect to all states, the second is Uhlmann's theorem stating that the fidelity is the maximum overlap between all purifications, and the last one is because of  Eq.~\eqref{app-eq:chi-norm}. This completes the proof of Eq.~\eqref{eq:purity-proof}.

    On the other hand, for orthogonality the following is obtained (below we drop $t'$ for convenience):
    \begin{align}
    \label{app-eq:ortho-preserving}
        D\left(\rho_{\upsilon},\rho_{\upsilon^\perp}\right)\geq&\  D\left(\chi_{\upsilon},\chi_{\upsilon^\perp}\right)-D\left(\rho_{\upsilon},\chi_{\upsilon}\right)-D\left(\rho_{\upsilon^\perp},\chi_{\upsilon^\perp}\right)
        \nonumber\\
       \geq &\ \sqrt{1-|\bra{\chi_\upsilon}\chi_{\upsilon^\perp}\rangle|^2}-\sqrt{1-F\left(\chi_\upsilon,\rho_{\upsilon}\right)}       \nonumber\\
       &\ -\sqrt{1-F\left(\chi_{\upsilon^\perp},\rho_{\upsilon^\perp}\right)}\nonumber\\
      \geq&\ 1-\frac{0.01(4\sqrt{\epsilon}+\epsilon^{0.45})}{1-(8\sqrt{\epsilon}+\epsilon^{0.45})} -2\sqrt{8\sqrt{\epsilon}+\epsilon^{0.45}}\nonumber\\
      =:&\ 1-h(\epsilon),
    \end{align}
    where we first used the triangle inequality, then the definition of the trace distance for pure states and its upper bound from Eq.~\eqref{eq:fidelity-distance_app}, and for the last inequality  Eqs.~\eqref{app-eq:arbit-chi1}-\eqref{app-eq:arbit-chi2},  \eqref{app-eq:chi-norm} and \eqref{app-eq:purity-proof} were employed. We also used the fact that through Eqs.~\eqref{app-eq:arbit-chi1}-\eqref{app-eq:arbit-chi2} and \eqref{app-eq:chi-norm}, one has
    \begin{equation}
        \sqrt{1-|\bra{\chi_\upsilon}\chi_{\upsilon^\perp}\rangle|^2}\geq1-0.01|\bra{\chi_\upsilon}\chi_{\upsilon^\perp}\rangle|
    \end{equation}
    for any $\epsilon\leq10^{-6}$.


\section{Proof of Proposition~\ref{thm:orthogonality_evolution}}
\label{app:orthogonality}
    To prove the proposition we will use the following technical lemma whose proof can be found in Appendix~\ref{app:lemma2}.
    \begin{lem}\label{lemma2}
    	 Let $\ket v \neq 0$ define a $d$-dimensional vector and  $A $ be a matrix of the same dimension. Consider the sequence $\ket v$, $A\ket v$, $A^2\ket v$, $A^3\ket v$, $\ldots$. Then, there exists
    	$n\leq d$ such that
    	\begin{equation}
    		A^n\ket v = \sum_{i=0}^{n-1} \lambda_i A^i\ket v
    	\end{equation}
     and the vectors $\ket{v}, \ldots, A^{n-1}\ket{v}$ are linearly independent. Moreover, it holds that
    \begin{equation}
    	\|\vec{\lambda}\|_1 \le n
    	\frac{(n+1)!}{2} \max \left[\|A\|_\infty,\|A\|_\infty^{n}\right],
    \end{equation}
    where $\|A\|_\infty$ is the operator norm of $A$.
    \end{lem}

   \begin{proof}[Proof of Proposition~\ref{thm:orthogonality_evolution}]

        We start by noticing that, through the data processing inequality, Eq.~\eqref{eq:two_orthogonal_states} implies
       \begin{equation}
           \forall t\leq t_f:\quad D\left(\e^{\L t}(\rho_1),\e^{\L t}(\rho_2)\right)\geq1-\epsilon.
       \end{equation}
       Applying the upper bound from Eq.~\eqref{eq:fidelity-distance_app} and the fact that for any two given states the fidelity is lower bounded by their Hilbert-Schmidt inner product, the above gives:
       \begin{equation}
       \label{eq:inner_product_bound}
           \forall t\leq t_f:\quad \Tr\left[\e^{\L t}(\rho_1)\e^{\L t}(\rho_2)\right]\leq\epsilon(2-\epsilon).
       \end{equation}

       Next, recall that for a quantum channel $\E$, there is a superoperator representation $\Phi_\E$ which is a $d^2$-dimensional matrix  defined by $\Phi_\E\Ket\rho=\Ket{\E(\rho)}$ where for a matrix $X=\sum x_{ij}\ketbra{i}{j}$ the vector $\Ket{X}=\sum x_{ij}\ket{ij}$ is its row-wise vectorised form.  Let now $\ket{v}:=\Ket{\rho_1}\Ket{\overline{\rho_2}}$,  $\E_t=\e^{\L t}$,  and $A_t:= \Phi_{\E_{t}} \otimes \overline{\Phi}_{\E_{t}}$ where overline denotes complex conjugation.
       Eq.~\eqref{eq:inner_product_bound} can then be restated as
       \begin{equation}
           \label{eq:vectorised_inner_product}
          \forall t\leq t_f:\quad \Bra{I}A_t\ket{v}\leq\epsilon(2-\epsilon),
       \end{equation}
       where $\Ket{I}$ is the vectorised form of $d^2$-dimensional identity matrix.
       Also, consider any $t_\star\in(0,t_f/(d^4-1)]$ and denote by $\ket{v'}:=\Ket{\sigma_1}\Ket{\overline{\sigma_2}}$  where for $i=1,2$
       \begin{equation}
            \sigma_i=\e^{\L\left(t_f-(d^4-1)t_\star\right)}(\rho_i).
        \end{equation}
       Therefore, Lemma~\ref{lemma2} implies that we can find a vector $\vec{\lambda} \in \R^{d^4}$
    	such that
    	\begin{equation}
    	\label{eq:vector_sigma}
    		A^{d^4}_{t_\star}\ket{v'} = \sum_{i=0}^{d^4-1} \lambda_i A_{t_\star}^i\ket{v'}
    	\end{equation}
    	and $\|\vec{\lambda}\|_1\leq H(d)$ with $H(d)$ given in Eq.~\eqref{eq:H(d)}. The latter is because of the Russo-Dye theorem stating that every positive linear map attains its norm at the identity, i.e., $\|\E\|_\infty=\|\E(\iden)\|_\infty$. Therefore, one straightforwardly gets $\|\Phi_\E\|_\infty\leq\sqrt{d}$ for any channel $\E$, where the bound is saturated when the map is a complete contraction into a pure state. Thus, we have $\|A_t\|_\infty = \|\Phi_{\E_t}\|_\infty^2\leq d$.
    	
    	Additionally, Eq.~\eqref{eq:vectorised_inner_product} implies that for any \mbox{$i\in\{0,\dots,d^{4}-1\}$}
    	\begin{equation}
    	    \Bra{I}A_{t_\star}^i\ket{v'}\leq\epsilon(2-\epsilon).
    	\end{equation}
    	This in turn means
    	\begin{equation}
    	    \Bra{I}A_{t_\star}^{d^4}\ket{v'}\leq H(d)\epsilon(2-\epsilon).
    	\end{equation}
    	Replacing $\sigma$ by $\rho$ and re-writing the above by matrix notation gives for all $t_\star\in(0,\frac{t_f}{d^4-1}]$
    	\begin{eqnarray}
    	    \!\!\!\Tr[\e^{\L(t_f+t_\star)}(\rho_1)\e^{\L(t_f+t_\star)}(\rho_2)]\leq H(d)\epsilon(2-\epsilon).
    	\end{eqnarray}
    	The latter should be applied as an updated assumption for $t \le t_f+\frac{t_f}{d^4-1}$ for the next interval, and so on.
        Thus, we get the bound in Eq.~\eqref{eq:inner_product_evolution}, which completes the proof.
    \end{proof}


\section{Proof of Lemma~\ref{lemma2}}
\label{app:lemma2}

\begin{proof}[Proof of Lemma~\ref{lemma2}]
    We note that the proof of the first part is obvious. To prove the bound, assume that $A^n\ket v = \sum_{i=0}^{n-1} \lambda_i
	A^i\ket v$ and $\ket v,\ldots,A^{n-1}\ket v$ are linearly independent, $\ket v$ is normalised to $1$ and $A\neq0$.

 In the first case, we consider $\|A\|_\infty = 1$.
	Define \mbox{$\ket{v_i} = A^{i-1}\ket v$} for $1 \le i \le n+1$. Let $\{\ket{w_i}\}$ for
	\mbox{$i=1,\ldots,n$}
	be an orthonormal set such that $\ket{w_1} =\ket{v_1}$ and $\ket{v_i} = \ket{z_i} + c_i \ket{w_i}$, where
	\mbox{$\ket{z_i} \in \mathrm{Span}\{\ket{v_1},\ldots,\ket{v_{i-1}}\}$}. We can
	express the action of $A$ in the basis $\ket{w_i}$, that is
	\begin{equation}
		A\ket{w_i} = \sum_{j=1}^{i+1} a_{j,i}\ket{w_j},
	\end{equation}
for $i<n$, while for $i=n$
\begin{equation}
	A\ket{w_n} = \sum_{j=1}^{n} a_{j,n}\ket{w_j},
\end{equation}
with $a_{2,1},\ldots,a_{n,n-1} \neq 0$.
We can write
\begin{equation}
	\ket{v_2} = A\ket{w_1} = a_{1,1}\ket{v_1} + a_{2,1}\ket{w_2}.
\end{equation}
By induction, assume that for all $k \le k_0$, we can write
\begin{equation*}
	\ket{v_{k+1}} = p_{k,1}\ket{v_1} + \ldots + p_{k,k}\ket{v_k} + a_{2,1}\cdots a_{k+1,k}\ket{w_{k+1}},
\end{equation*}
where $p_{k,i}$ are polynomials of variables $(a_{j,i})$ satisfying
\begin{equation}
	p_{k,i} = \sum_{Z_k} \pm a_{j_1,i_1} \cdots a_{j_r,i_r}
\end{equation}
for $|Z_k| \le f(k)$. In particular, for $p_{1,1}$ we have $f(1)
= 1$. By assumption, for $i = 2,\ldots,k_0+1$ we have
\begin{equation}
	\ket{w_i} = \frac{\ket{v_i} - p_{i-1,1}\ket{v_1}- \ldots -
	p_{i-1,i-1}\ket{v_{i-1}}}{a_{2,1} \cdots a_{i,i-1}}.
\end{equation}
Then,
\begin{align}
		\ket{v_{k_0+2}}&= A\ket{v_{k_0+1}}= p_{k_0,1}\ket{v_2} + \ldots + p_{k_0,k_0}\ket{v_{k_0+1}}
		\nonumber\\
  &\qquad\qquad\qquad+
		a_{2,1}\cdots a_{k_0+1,k_0}A\ket{w_{k_0+1}}\nonumber\\
		&=p_{k_0,1}\ket{v_2} + \ldots + p_{k_0,k_0}\ket{v_{k_0+1}}
    	\nonumber\\&\quad+
		a_{2,1}\cdots a_{k_0+1,k_0}\sum_{j=1}^{k_0+2}
		a_{j,k_0+1}\ket{w_j}\nonumber\\
	    &\coloneqq p_{k_0+1,1}\ket{v_1} + \ldots + p_{k_0+1,k_0+1}\ket{v_{k_0+1}}\nonumber
		\\&\quad+
		a_{2,1}\cdots a_{k_0+2,k_0+1}\ket{w_{k_0+2}}.
\end{align}
Notice that $f(k_0 + 1) \le (k_0+2)f(k_0)$.
That means we have $f(k) \le (k+1)!/2$.
We continue the induction until we express $\ket{v_n}$ in the same format and then we
can express $A^n\ket{v}$ as
\begin{equation}
	\begin{split}
		\ket{v_{n+1}} &= p_{n-1,1}\ket{v_2} + \ldots + p_{n-1,n-1}\ket{v_{n}}
    	\\&\quad +
		a_{2,1}\cdots a_{n,n-1}A\ket{w_{n}}\\
		&\coloneqq p_{n,1}\ket{v_1} + \ldots + p_{n, n}\ket{v_{n}}.
	\end{split}
\end{equation}
Eventually,
\begin{align}
	\|\vec{\lambda}\|_1 = \sum_{k=1}^n|p_{n,k}| \le n  f(n) \le n 	\frac{(n+1)!}{2} .
\end{align}

In the second case, we consider arbitrary $A \neq 0$. Then,
\begin{align}
   \!\! \|A\|_\infty^n\! \left(\dfrac{A}{\|A\|_\infty}\right)^n\!\!\!\ket v &= A^n\ket v = \sum_{i=0}^{n-1} \lambda_i 	A^i\ket v \nonumber\\ &= \sum_{i=0}^{n-1} \lambda_i \|A\|_\infty^i 	\left(A/\|A\|_\infty\right)^i\ket v.
\end{align}
According to the first case, $\sum_{i=0}^{n-1} |\lambda_i| \|A\|_\infty^{i-n} \le  n 	\frac{(n+1)!}{2}$, which implies $\|\vec{\lambda}\|_1 \le n
    	\frac{(n+1)!}{2} \max \left[\|A\|_\infty,\|A\|_\infty^{n}\right]$.
\end{proof}


\bibliographystyle{quantum}
\bibliography{Embedding}

\end{document}